\documentclass[11pt,a4paper]{article}

\pdfoutput=1

\usepackage{jinstpub}
\usepackage[english]{babel}
\usepackage{subfigure}
\usepackage{xspace}
\usepackage{booktabs}
\usepackage{textcomp}
\usepackage{lineno}

\newcommand{\mr}[1]{\mathrm{#1}}

\def\EeV{\ifmmode {\mathrm{\ Ee\kern -0.1em V}}\else
                   \textrm{Ee\kern -0.1em V}\fi\xspace}%
\def\PeV{\ifmmode {\mathrm{\ Pe\kern -0.1em V}}\else
                   \textrm{Pe\kern -0.1em V}\fi\xspace}%
\def\TeV{\ifmmode {\mathrm{\ Te\kern -0.1em V}}\else
                   \textrm{Te\kern -0.1em V}\fi\xspace}%
\def\MeV{\ifmmode {\mathrm{\ Me\kern -0.1em V}}\else
                   \textrm{Me\kern -0.1em V}\fi\xspace}%
\def\GeV{\ifmmode {\mathrm{\ Ge\kern -0.1em V}}\else
                   \textrm{Ge\kern -0.1em V}\fi\xspace}%
\def\keV{\ifmmode {\mathrm{\ ke\kern -0.1em V}}\else
                   \textrm{ke\kern -0.1em V}\fi\xspace}%
\def\MeV{\ifmmode {\mathrm{\ Me\kern -0.1em V}}\else
                   \textrm{Me\kern -0.1em V}\fi\xspace}%
\def\eV{\ifmmode {\mathrm{\ e\kern -0.1em V}}\else
                   \textrm{e\kern -0.1em V}\fi\xspace}%

\def \mal      {Malarg\"{u}e\xspace}
\def \pao      {Pierre Auger Observatory\xspace}
\def \rnd      {R\&D\xspace}

\def \degree   {$^{\circ}$\xspace}
\def \degC     {$^{\circ}$C\xspace}
\def \inch     {$^{\prime\prime}$\xspace}
\def \vaod     {$\tau_\mr{aer}$\xspace}
\def \lidar    {\textsc{Lidar}\xspace}
\def \lidars   {\textsc{Lidar}s\xspace}
\def \us       {$\text{\textmu s}$\xspace}

\def \n2       {N$_2$\xspace}
\def \o2       {O$_2$\xspace}
\def \h2o      {H$_2$O\xspace}
\def \d        {\text{d}}
\def \zemax    {Zemax\textregistered{}\xspace}

\title{Joint Elastic Side-Scattering \lidar and Raman \lidar Measurements of Aerosol Optical Properties in South East Colorado}

\author[a]{L.~Wiencke,}
\author[b]{V.~Rizi,}
\author[a,c,1]{M.~Will,
  \note{now at Instituto de Astrof\'isica de Canarias, La Laguna, Tenerife, Spain}}
\author[a]{C.~Allen,}
\author[a]{A.~Botts,}
\author[a]{M.~Calhoun,}
\author[a]{B.~Carande,}
\author[a]{J.~Claus,}
\author[a]{M.~Coco,}
\author[a]{L.~Emmert,}
\author[a]{S.~Esquibel,}
\author[b]{A.~F.~Grillo,}
\author[a]{L.~Hamilton,}
\author[a]{T.~J.~Heid,}
\author[b]{M.~Iarlori,}
\author[c]{H.-O.~Klages,}
\author[d]{M.~Kleifges,}
\author[a]{B.~Knoll,}
\author[a]{J.~Koop,}
\author[c]{H.-J.~Mathes,}
\author[d]{A.~Menshikov,}
\author[a]{S.~Morgan,}
\author[a]{L.~Patterson,}
\author[b]{S.~Petrera,}
\author[a]{S.~Robinson,}
\author[a]{C.~Runyan,}
\author[a]{J.~Sherman,}
\author[a]{D.~Starbuck,}
\author[a]{M.~Wakin}
\author[a]{and O.~Wolf}

\affiliation[a]{Colorado School of Mines, Golden, Colorado, USA}
\affiliation[b]{INFN/LNGS and CETEMPS, Universit\`{a} Degli Studi dell'Aquila, L'Aquila, Italy}
\affiliation[c]{Karlsruher Institut f\"{u}r Technologie, Institut f\"{u}r Kernphysik (IKP), Karlsruhe, Germany}
\affiliation[d]{Karlsruher Institut f\"{u}r Technologie, Institut f\"{u}r Prozessdatenverarbeitung und Elektronik (IPE), Karlsruhe, Germany}

\emailAdd{lwiencke@mines.edu, vincenzo.rizi@aquila.infn.it, martin.will@ifae.es}

\abstract{
    We describe an experiment, located in south-east Colorado, USA, that
    measured aerosol optical depth profiles using two \lidar techniques. Two
    independent detectors measured scattered light from a vertical UV laser
    beam. One detector, located at the laser site, measured light via the
    inelastic Raman backscattering process. This is a common method used in
    atmospheric science for measuring aerosol optical depth profiles. The other
    detector, located approximately 40\,km distant, viewed the laser beam from
    the side. This detector featured a 3.5\,m$^2$ mirror and measured
    elastically scattered light in a bistatic \lidar configuration following the
    method used at the Pierre Auger cosmic ray observatory. The goal of this
    experiment was to assess and improve methods to measure atmospheric clarity,
    specifically aerosol optical depth profiles, for cosmic ray UV fluorescence
    detectors that use the atmosphere as a giant calorimeter. The experiment
    collected data from September 2010 to July 2011 under varying conditions of
    aerosol loading. We describe the instruments and techniques and compare the
    aerosol optical depth profiles measured by the Raman and bistatic \lidar
    detectors.
}


\begin{document}

\maketitle


\section{Introduction
\label{sec:introduction}}

\begin{figure}[!t]
  \centering
  \includegraphics[width=.75\linewidth,clip]{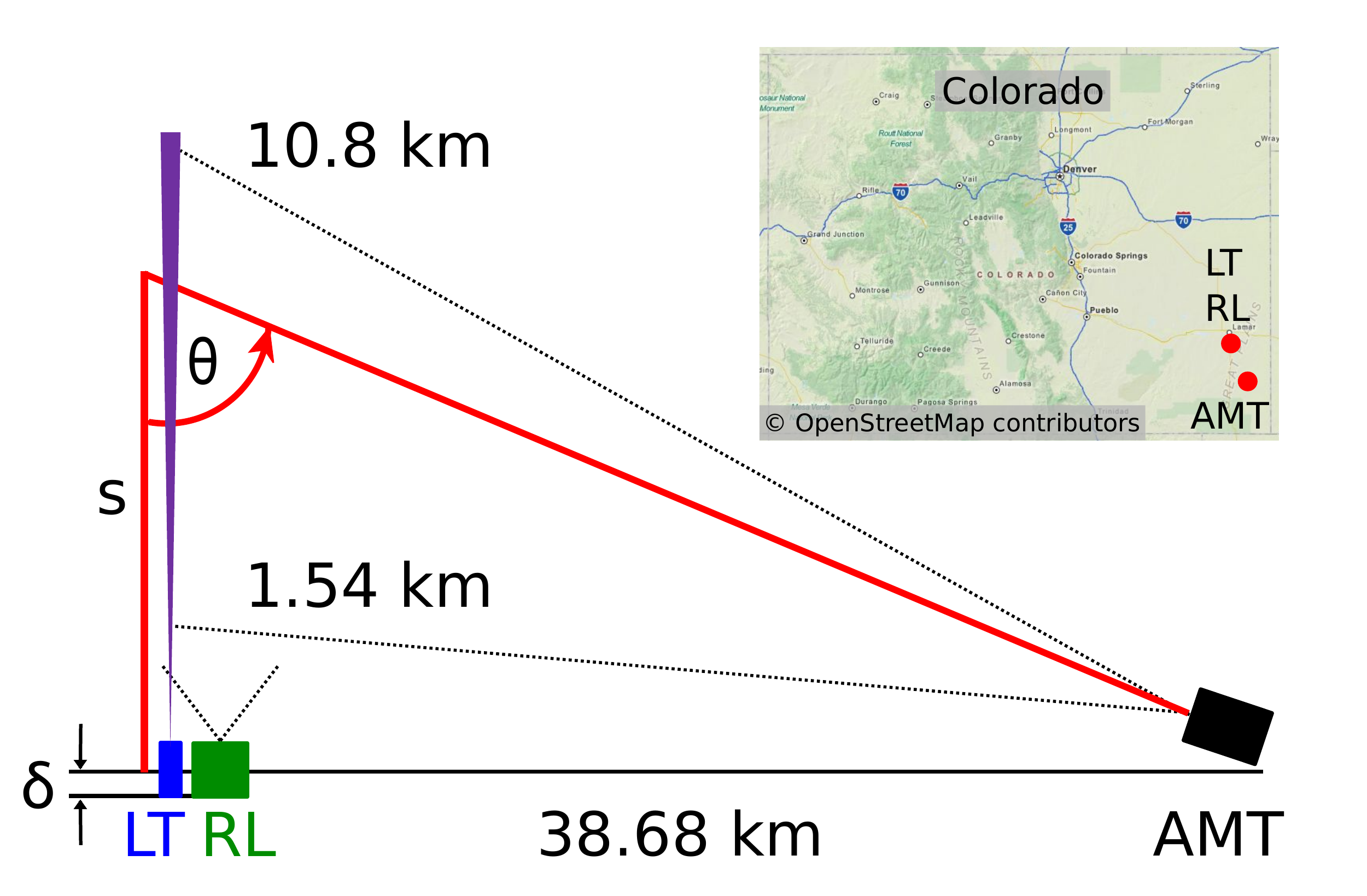}
  \caption{
    \label{fig:geodiagram}
    Geometrical arrangement, viewed from the side, of the three instruments used
    in this experiment. A pulsed UV laser (LT) directed a vertical beam into the
    atmosphere. Scattered light was measured by the Raman \lidar Detector (RL),
    and by the Atmospheric Monitoring Telescope (AMT). A map of
    Colorado~\cite{OSM} showing the location of the instruments is inserted.
  }
\end{figure}

Understanding certain properties of the troposphere is critical for ultra high
energy cosmic ray observatories that use the atmosphere as an enormous
calorimeter. An accurate air density profile is needed to measure the energy
deposited in the atmosphere from a cosmic ray extensive air shower (EAS). The
amount of UV fluorescence light generated by an EAS is proportional to the
energy deposited. The optical clarity of the atmosphere affects the amount of
light that reaches the air fluorescence detector (FD)~\cite{Abraham:2009fd}. The
latter is the calibration factor with the largest and fastest variation with
time.

The most important component of optical clarity is the vertical aerosol optical
depth profile, \vaod. Although usually much smaller in magnitude than the
corresponding optical depth profile of the relatively stable molecular
component, \vaod can vary significantly on the time scale of hours. Under hazy
conditions, the apparent brightness of the highest energy EASs can decrease by a
factor of 2 or more. Hourly \vaod measurements are required. Compounding the
measurement challenge is that an FD requires dark, quiet skies to measure EASs
effectively. The quiet sky requirement has precluded routine operation of Raman
\lidars. To accumulate a statistically significant return profile, many laser
pulses (of order $10^5$) must be fired into the sky. These pulses will also
trigger the FD and cause too much dead time. Still, Raman scattering by \n2
molecules provides a known ``atmospheric mirror'' that does not depend on the
quantity (\vaod) to be measured. Hence it is the preferred method in atmospheric
sciences for measuring aerosol profiles.

To evade the apparent choice between measuring the atmosphere well or measuring
the cosmic rays well, the High Resolution Fly's Eye~\cite{Abbasi:2006aer} and
the Pierre Auger Observatory~\cite{Abreu:2013clf, Abraham:2010atmo} adapted an
elastic bistatic \lidar technique to measure \vaod. A UV pulsed laser with
optics that aim the beam vertically is operated in the field of view of the FD.
The laser wavelength falls near the middle of the EAS UV fluorescence spectrum.
As a laser pulse propagates upward through the atmosphere, the elastically
scattered light produces a track in the FD which also records tracks of
fluorescence light from EASs. The energy of the laser is set to approximate the
amount of light observed, under clear conditions, from the very rare highest
energy EASs. The laser is placed at a distance from the FD that is typical of
the the highest energy EASs observed (10--30\,km). With this bistatic technique,
the FD provides the receiver and \vaod can be measured every 15~minutes with
about 200~laser shots per hour.

In this paper we describe an atmospheric research and development program (\rnd)
that made the first comparison between this elastic bistatic technique and the
traditional backscatter Raman \lidar technique. The program was conducted in
south-eastern Colorado which is a candidate site for a possible giant cosmic ray
observatory. The Raman \lidar receiver developed for this \rnd would later be
installed at the Pierre Auger Observatory central laser
facility~\cite{Fick:2006}. The arrangements of instruments used in the Colorado
\rnd (Fig.~\ref{fig:geodiagram}) included a frequency-tripled Nd:YAG laser that
generated a vertical pulsed beam, a collocated Raman \lidar receiver, and a
simplified FD telescope, the Atmospheric Monitoring Telescope (AMT), located
about 39\,km distant.

This paper is organized as follows. Raman \lidar and side-scattering detector
theory is summarized in Sec.~\ref{sec:theory}. Technical details and system
performances are presented in Sec.~\ref{sec:instruments}. Mode of operations and
examples of observations are described in Sec.~\ref{sec:operations} and
~\ref{sec:results}, respectively. Finally Sec.~\ref{sec:conclusion} includes a
summary and discussion about future applications of the techniques.

\section{Raman LIDAR and Side-Scattering Theory: Measurements of Atmospheric Aerosol Optical Properties
\label{sec:theory}}

The measurement of atmospheric aerosol optical properties are based on the
well-known single scattering \lidar equation and the Raman \lidar principle. In
our experiment, the laser has a negligibly small bandwidth of about
1\,cm$^{-1}$, and the beam is linearly polarized for Raman \lidar (RL) and
unpolarized for side-scattering measurements with the AMT. If $L_0$ is the
number of photons emitted per laser pulse at wavelength $\lambda_0$~=~354.7\,nm,
the number of photons reaching the height $s = ct / 2$ (where $c$ is the speed
of light, and $t$ the measured time-of-flight of the photons from emission to
height $s$ and back to the receiver) above the laser transmitter (LT) is
\begin{linenomath*}
\begin{equation*}
  L(s)=L_0 \cdot T_{\rm aer}(s) \cdot T_{\rm mol}(s),
\end{equation*}
\end{linenomath*}
where $T_{\rm aer}(s)$ and $T_{\rm mol}(s)$ are the transmission factors that
account for the optical extinction due to aerosol (Mie scattering) and air
molecules (Rayleigh scattering),
\begin{linenomath*}
\begin{align*}
  T_{\rm aer}(s) &= \exp \left( -\int_0^s \! \alpha_{\rm aer}(s') \, \d s' \right), \\
  T_{\rm mol}(s) &= \exp \left( -\int_0^s \! \alpha_{\rm mol}(s') \, \d s' \right).
\end{align*}
\end{linenomath*}
$\alpha_{\rm aer}(s)$ and $\alpha_{\rm mol}(s)$ are the aerosol and molecular
extinction coefficients,
\begin{linenomath*}
\begin{align*}
  \alpha_{\rm aer}(s) &= \int_0^\infty \! Q_{\rm ext}(r,m,\lambda_0)\, n_{\rm aer}(s,r)\, \pi r^2\, \d r, \\
  \alpha_{\rm mol}(s) &= \sigma_{\rm mol}(s) \cdot n_{\rm mol}(s),
\end{align*}
\end{linenomath*}
where $n_{\rm aer}(s,r)$ is the aerosol size distribution, $Q_{\rm
ext}(r,m,\lambda_0)$ the Mie extinction efficiency of an aerosol particle of
radius $r$, and $m$ the index of refraction. $\sigma_{\rm mol}$ is the total
Rayleigh scattering cross section at $\lambda_0$, and $n_{\rm mol}(s)$ is the
number density of air molecules.

A fraction of the $L(s)$ photons is elastically backscattered (scattering angle
$\pi$) to the RL by the air molecules and the aerosols,
\begin{linenomath*}
\begin{equation*}
  L_{\rm E}(s)=L(s) \cdot \left[ \beta_{\rm mol}(s, \pi) + \beta_{\rm aer}(s, \pi)\right]
    \cdot \Delta s \cdot T_{\rm aer}(s) \cdot T_{\rm mol}(s) \cdot \frac{A_{\rm E}}{s^2} \cdot G_{\rm E}(s).
\end{equation*}
\end{linenomath*}
These are the detected photons if there is no discrimination of the polarization
state. The spectral width of the detector is large enough (i.\,e., it has the
same collection efficiency over a wavelength range of few nanometers centered
around $\lambda_0$) to detect the unshifted and partly depolarized Cabannes line
and the wavelength-shifted and fully depolarized pure rotational Raman lines on
both sides of the Cabannes line, coming mainly from the \n2 and \o2 molecules
(contributing at a level of about 2\% of the total Rayleigh scattering).
$A_{\rm E}$ is a constant and is proportional to the area of the RL telescope and the
overall detection efficiency of the corresponding \lidar channel, $\Delta s$ is
the vertical resolution, which depends on the detector's electronics. $G_{\rm
E}(s)$ is the geometrical overlap function of the elastic \lidar channel, a
measure of the height-dependent collecting efficiency of the receiver telescope.
It depends on the laser divergence, the receiver field of view and on the
distance between telescope and laser axis.
$\beta_{\rm mol}(s, \pi)$ is the molecular backscattering coefficient,
\begin{linenomath*}
\begin{equation*}
  \beta_{\rm mol}(s, \pi)=\frac{\d \sigma_{\rm mol}(s, \pi)}{\d \Omega} \cdot n_{\rm mol}(s),
\end{equation*}
\end{linenomath*}
depending on the local atmospheric molecular number density, $n_{\rm mol}(s)$, and the differential Rayleigh backscattering cross section, $\frac{\d \sigma_{\rm mol}(s, \pi)}{\d \Omega}$, at $\lambda_0$. $\beta_{\rm aer}(s, \pi)$
is the aerosol backscatter coefficient,
\begin{linenomath*}
\begin{equation*}
  \beta_{\rm aer}(s, \pi)=\int_0^\infty \! Q_{\rm bck}(r,m,\lambda_0)\, n_{\rm aer}(s,r)\, \pi r^2 \, \d r,
\end{equation*}
\end{linenomath*}
where $Q_{\rm bck}(r,m,\lambda_0)$ is the Mie backscatter efficiency. 

The Raman-backscattered \lidar return (i.\,e., by \n2 molecules) for a
polarization-independent observation of the complete backscatter signal can be
written as
\begin{linenomath*}
\begin{equation*}
  L_{\rm R}(s)=L(s) \cdot \left[ \beta_{{\rm N}_{2}}^{\rm R}(s, \pi) \right] \cdot \Delta s \cdot T_{\rm aer}^{\rm R}(s)
    \cdot T_{\rm mol}^{\rm R}(s) \cdot \frac{A_{\rm R}}{s^2} \cdot G_{\rm R}(s).
\end{equation*}
\end{linenomath*}
The \n2 Raman scattered photons have a shifted (central) wavelength
$\lambda_{\rm R}$ of 386.7\,nm, if the exciting wavelength is 354.7\,nm. The
Raman backscatter coefficient at wavelength $\lambda_{\rm R}$, $\beta_{{\rm
N}_{2}}^{\rm R}(s, \pi)$, is the product of the differential \n2 Raman
backscattering cross section and the molecular number density of the \n2
molecules at height $s$, $n_{{\rm N}_{2}}(s)$,
\begin{linenomath*}
\begin{equation*}
  \beta_{{\rm N}_{2}}^{\rm R}(s, \pi)= \frac{\d \sigma_{{\rm N}_{2}}^{\rm R}(\pi)}{\d \Omega} \cdot n_{{\rm N}_{2}}(s).
\end{equation*}
\end{linenomath*}
$\d \sigma_{{\rm N}_{2}}^{\rm R}(\pi)/\d \Omega$ is the Raman backscattering
differential cross section of the Stokes vibration-rotation Raman lines: the
central Q branch and the O and S side-branches that span about 5\,nm around the
central wavelength. The O and S branches contribute about 14\% to the total
differential Raman backscattering cross section. $T_{\rm aer}^{\rm R}(s)$ and
$T_{\rm mol}^{\rm R}(s)$ are the aerosol and molecular optical transmissions at
$\lambda_{\rm R}$, $A_{\rm R}$ is proportional to the area of the RL telescope,
and to the detection efficiency of the \n2 Raman \lidar channel. $G_{\rm R}(s)$
is the geometrical overlap function of the Raman \lidar channel. Note that
$G_{\rm R}(s) \neq G_{\rm E}(s)$ is possible because, in principle, the field of
view of the \lidar receiver can be different for each of the \lidar channels.

A similar equation can be written for the Raman-backscattered \lidar return by
water vapor molecules, the \h2o vibration-rotation Raman lines fall at 407.5\,nm
(Q branch), and the O and S branches (covering an 8\,nm interval around the
central wavelength) contribute about 9\% to the total differential Raman
backscattering cross section. The capability of our RL detector of measuring the
water vapor content is not reported in this paper, the main reasons are
discussed in Sec.~\ref{sec:theoryRL}.

$L_{\rm S}(s)$ are the photons side-scattered towards the AMT from a height $s$, the
scattering angle is $(\pi - \theta)$ (see Fig.~\ref{fig:geodiagram})
\begin{linenomath*}
\begin{equation*}
  L_{\rm S}(s)=L(s) \cdot \left[ \beta_{\rm mol}(s, \pi - \theta ) + \beta_{\rm aer}(s, \pi - \theta ) \right]
    \cdot \Delta s \cdot \left[ T_{\rm aer}(s) \cdot T_{\rm mol}(s)\right]^{\sec(\theta)} \cdot \frac{A_{\rm S} \cos^2(\theta)}{s^2},
\end{equation*}
\end{linenomath*}
where $A_{\rm S}$ accounts for the aperture and the detection efficiency of the
AMT. The scattering coefficients can be written as
\begin{linenomath*}
\begin{align}
  \label{eq:beta_pi-theta}
  \beta_{\rm mol}(s, \pi - \theta ) &= \beta_{\rm mol}(s, \pi) \cdot \frac{P_{\rm mol}(\pi - \theta)}{P_{\rm mol}(\pi)}, \nonumber \\
  \beta_{\rm aer}(s, \pi - \theta ) &= \beta_{\rm aer}(s, \pi) \cdot \frac{P_{\rm aer}(\pi - \theta)}{P_{\rm aer}(\pi)}.
\end{align}
\end{linenomath*}
Here, $P_{\rm mol}(\phi)$ is the Rayleigh scattering phase function, i.\,e., the
probability of a photon being scattered in the direction $\phi$, and $P_{\rm
aer}(\phi)$ is the aerosol scattering phase function.

\subsection{Raman \lidar
\label{sec:theoryRL}}

The Raman \lidar is designed to measure the vertical profiles of the aerosol
backscatter coefficients $\beta_{\rm aer}(s, \pi)$ and extinction coefficients
$\alpha_{\rm aer}(s)$ at the laser wavelength $\lambda_0$. The design of the RL
receiver includes the capability to detect the \h2o Raman backscattered photons,
and with this to measure the water vapor vertical profile. After the first
testing phase of the RL, it was found that the liquid light guide (see below) of
the receiver shows a fluorescent re-emission when transporting the \lidar
backscattered photons, and the fluorescence signal has a spectral signature that
directly superimposes the \h2o Raman backscatter signal.

Rayleigh/Mie and Raman \lidar inversion methods for the estimation of the
aerosol optical properties are well known, and their combination leads to an
improvement of the results~\cite{Pappalardo}. The aerosol extinction can be
determined from \n2 Raman \lidar return, through the application of the
expression
\begin{linenomath*}
\begin{equation}
  \label{eq:alpharaman}
  \alpha_{\rm aer}(s)= \frac{\frac{s^2 L_{\rm R}(s)}{n_{{\rm N}_{2}}(s)} \cdot \frac{\d}{\d s} \left[ \frac{n_{{\rm N}_{2}}(s)}{s^2 L_{\rm R}(s)}\right]
  -\alpha_{\rm mol}(s)-\alpha_{\rm mol}^{\rm R}(s)}{1+\left(\frac{\lambda_0}{\lambda_{\rm R}}\right)^k},
\end{equation}
\end{linenomath*}
where $\alpha_{\rm mol}(s)$ and $\alpha_{\rm mol}^{\rm R}(s)$ are the molecular
extinctions at $\lambda_0$ and $\lambda_{\rm R}$, respectively. $n_{{\rm
N}_{2}}(s)$ is the number density of \n2 molecules. The wavelength scaling of
aerosol extinction is proportional to $\lambda^k$, where $k$ is the \r{A}ngstrom
coefficient which is in general a function of altitude, since it depends on
aerosol properties. A good assumption is to set a value of the \r{A}ngstrom 
coefficient in the interval $k = 1.0 \pm 0.5$~\cite{Mueller:Raman}; this marginally 
influences the systematic uncertainty affecting the aerosol extinction coefficient, 
i.e., it introduces an error on the aerosol optical depth less than few 
percentages. To estimate the aerosol extinction, the derivative of a function
containing the \lidar return has to be calculated, represented in discrete range
bins. The numerical technique to accomplish this calculation can be the sliding
linear least-squares fit. Both $\alpha_{\rm aer}(s)$ and its uncertainty could
be misevaluated if data acquisition and analysis are not correctly accomplished.

The uncertainties affecting $\alpha_{\rm aer}(s)$ are the statistical
uncertainty due to signal detection, the systematic uncertainty associated with
the estimation of the molecular number density and the Rayleigh scattering cross
section, the systematic uncertainty associated with the evaluation of the
aerosol scattering wavelength dependence (\r{A}ngstrom coefficient), and the
uncertainties introduced by operational procedures such as signal averaging
(accumulating \lidar returns), and by applying, for example, derivative digital
filters.

An additional systematic uncertainty that should be accounted for is due to the
geometrical overlap function of the \lidar. In a range of heights where the
optical overlap between the laser and the field of view of the receiving mirror
is range dependent, this uncertainty can be quite important. The atmospheric
temperature and pressure profiles from balloon soundings or global
meteorological models (i.\,e., GDAS~\cite{GDASinformation}) are used to estimate
the Rayleigh scattering components of the \n2 Raman \lidar return.

Another quantity that is usually evaluated is the vertical aerosol optical depth
\vaod. The \vaod profile between the range heights $s_{\rm l}$ and $s$ is
defined as
\begin{linenomath*}
\begin{equation}
  \tau_{\rm aer}(s_{\rm l},s)=\int_{s_{\rm l}}^{s} \alpha_{\rm aer}(s')\,\d s'.
\end{equation}
\end{linenomath*}
Typically, $s_{\rm l}$ coincides with the lower most height at which the \n2
Raman \lidar return is not affected by the optical overlap distortions, the
upper limit $s$ is in the free troposphere. Below $s_{\rm l}$ the extinction
coefficient is assumed to be a constant $\alpha_{\rm aer}(s_{\rm l})$. The total
\vaod becomes
\begin{linenomath*}
\begin{equation}
  \tau_{\rm aer}(s)=\alpha_{\rm aer}(s_{\rm l}) \cdot s_{\rm l} + \tau_{\rm aer}(s_{\rm l},s).
\end{equation}
\end{linenomath*}

Alternatively, the evaluation of the \vaod can be done directly from the \n2
Raman \lidar return,
\begin{linenomath*}
\begin{equation}
  \tau_{\rm aer}(s_{\rm l},s)=- \frac{\log \left(\frac{L_{\rm R}(s) \cdot s^{2} \cdot C}{T_{\rm mol}^{\rm R}(s)
  \cdot T_{\rm mol}(s) \cdot n_{\rm mol}(s)} \right)}{1+\left(\frac{\lambda_0}{\lambda_{\rm R}}\right)^k} .
\end{equation}
\end{linenomath*}
The constant $C$ is determined imposing that \vaod below $s_{\rm l}$ is a linear
function through the origin of the range $s$. Again, this means that the aerosol
extinction coefficient is assumed constant in the height range below $s_{\rm
l}$.

The aerosol volume backscattering coefficient is evaluated starting from the
ratio between the elastic and Raman \lidar returns,
\begin{linenomath*}
\begin{equation}
  \label{eq:betaraman}
  \beta_{\rm aer}(s)=\beta_{\rm mol}(s)\cdot \left[ 0.781 \cdot \frac{L_{\rm E}(s)}{L_{\rm R}(s)} \cdot \frac{A_{\rm R} \cdot G_{\rm R}(s)}{A_{\rm E} \cdot G_{\rm E}(s)}
    \cdot \frac{\frac{\d \sigma_{\rm N_{2}}^{\rm R}(\pi)}{\d \Omega}}{\frac{\d \sigma_{\rm mol}(s, \pi)}{\d \Omega}}
    \cdot \frac{T_{\rm mol}^{\rm R}(s) \, T_{\rm aer}^{\rm R}(s)}{T_{\rm mol}(s) \, T_{\rm aer}(s)} - 1 \right]
\end{equation}
\end{linenomath*}

The design of our Raman \lidar receiver (the telescope is coupled to the
detector box through a liquid light guide) assigns the same optical overlap
modulation to the Rayleigh/Mie elastic and inelastic \n2 Raman \lidar channels.
Since the evaluation of $\beta^{\lambda_0}_{\rm aer}(s)$ involves the ratio
between these two \lidar returns, the estimation of the aerosol backscattering
coefficient results are independent of the \lidar geometrical overlap, assuming
$G_{\rm R}(s) = G_{\rm E}(s)$. After the estimation and the removal of the
Rayleigh scattering and backscattering contributions and of the aerosol
transmission as evaluated from the aerosol extinction, a calibration is needed.
In other words, the quantity
\begin{linenomath*}
\begin{equation}
  \label{eq:AR_AE}
  \frac{A_{\rm R}}{A_{\rm E}} \cdot \frac{\frac{\d \sigma_{\rm N_{2}}^{\rm R}(\pi)}{\d \Omega}}
    {\frac{\d \sigma_{\rm mol}(s, \pi)}{\d \Omega}}
\end{equation}
\end{linenomath*}
has to be estimated. Usually this is done by imposing $\beta^{\lambda_0}_{\rm
aer}(s) = 0$ in a range of altitudes free of aerosols, i.\,e., in the upper
troposphere~\cite{Bockmann}.

The uncertainties affecting $\beta^{\lambda_0}_{\rm aer}(s)$ are mainly due to
the statistical uncertainty in the signal detection, the systematic uncertainty
associated with the estimation of the molecular number density (i.\,e., from
pressure and temperature vertical profiles) and the Rayleigh scattering cross
section, and the uncertainties introduced by operational (retrieval) procedures,
as the estimation of the quantity in equation~\eqref{eq:AR_AE}.

\subsection{Side-scattering experiment}

The AMT is designed to measure \vaod directly. The method relies on the
assumption that during a measurement as close in time as possible to the actual
sampling of the atmosphere $L_{\rm S}(s)$, conditions can be found in which the
atmosphere can be considered free of aerosols. In this case the side-scattering
return can be expressed as
\begin{linenomath*}
\begin{equation*}
  L_{\rm S}^{\rm clean}(s) = L_0 \cdot T_{\rm mol}(s) \cdot \left[ \beta_{\rm mol}(s, \pi - \theta ) \right]
    \cdot \Delta s \cdot \left[T_{\rm mol}(s)\right]^{\sec(\theta)} \cdot \frac{A_{\rm S} \cos^2(\theta)}{s^2}.
\end{equation*}
\end{linenomath*}
The ratio between $L_{\rm S}(s)$ and $L_{\rm S}^{\rm clean}(s)$ is
\begin{linenomath*}
\begin{equation*}
  \frac{L_{\rm S}(s)}{L_{\rm S}^{\rm clean}(s)} = \left[T_{\rm aer}(s)\right]^{1+\sec(\theta)}
    \cdot \frac{\left[ \beta_{\rm mol}(s, \pi - \theta ) + \beta_{\rm aer}(s, \pi - \theta ) \right]}
    {\left[ \beta_{\rm mol}(s, \pi - \theta ) \right]}.
\end{equation*}
\end{linenomath*}
Recalling Eq.~\eqref{eq:beta_pi-theta}, considering that in our configuration
$\theta \in \left[74.8^{\circ}, 87.8^{\circ} \right]$, and that for most of the
atmospheric aerosol the scattering phase function is peaked in forward and
backward directions, it can be assumed that~\cite{Kokhanovsky:1998}
\begin{linenomath*}
\begin{equation*}
  P_{\rm aer}(\pi - \theta) \ll P_{\rm aer}(\pi).
\end{equation*}
\end{linenomath*}
Because the laser light is unpolarized and $\frac{P_{\rm mol}(\pi -
\theta)}{P_{\rm mol}(\pi)} \gtrsim 0.5$,
\begin{linenomath*}
\begin{equation*}
  \beta_{\rm mol}(s, \pi - \theta ) \gg \beta_{\rm aer}(s, \pi - \theta ).
\end{equation*}
\end{linenomath*}
Finally, \vaod can be written as
\begin{linenomath*}
\begin{equation}
  \label{eq:datanorm}
  \tau_{\rm aer}(s) = -\frac{1}{1+\sec(\theta) } \cdot \log \left[ \frac{L_{\rm S}(s)}{L_{\rm S}^{\rm clean}(s)} \right]
\end{equation}
\end{linenomath*}
The errors on \vaod are mainly due to the relative calibration of the AMT, the
relative uncertainty in the determination of the $L_{\rm S}^{\rm clean}(s)$, and
the statistical fluctuations introduced by signal averaging.

\section{Description and Performances of the Instruments
\label{sec:instruments}}

The main criteria that define the constraints of our experimental setup are
\begin{itemize}
  \item nighttime measurements in new and crescent moon phases,
  \item high accuracy measurements of aerosol optical properties in
    the planetary boundary layer and in the lower troposphere,
  \item remote operations and minimal maintenance.
\end{itemize}
The laser transmitter system (LT), collocated Raman \lidar (RL) detector and
distant side-scattering detector called Atmospheric Monitoring Telescope (AMT)
comprise the experiment (Fig.~\ref{fig:geodiagram}). The LT and RL point
vertically. The optical axis of the AMT is pointed above the LT, the field of
view extends from 2.2\degree to 15.2\degree. The corresponding sampling heights
at the LT location range from 1.5\,km to 10.6\,km above ground level. The LT and
RL are located at 37.9228\degree N, 102.6109\degree W and 1198\,m a.s.l.\ in Prowers County,
Colorado, about 15\,km south of Lamar. The AMT is located at 37.6010\degree N, 102.4420\degree W, 1279\,m a.s.l.\ close to the town of Two Buttes in Baca County,
Colorado. The sites are 38.7\,km apart. The difference in altitude between the
two sites $\delta$ is 81\,m, this difference as well as the curvature of the
Earth are taken into account in the data analysis. In the azimuth direction, the
AMT points 337.52\degree clockwise from north.

\subsection{The Laser Transmitter}

\begin{figure}[t]
  \begin{center}
    \includegraphics[width=12cm]{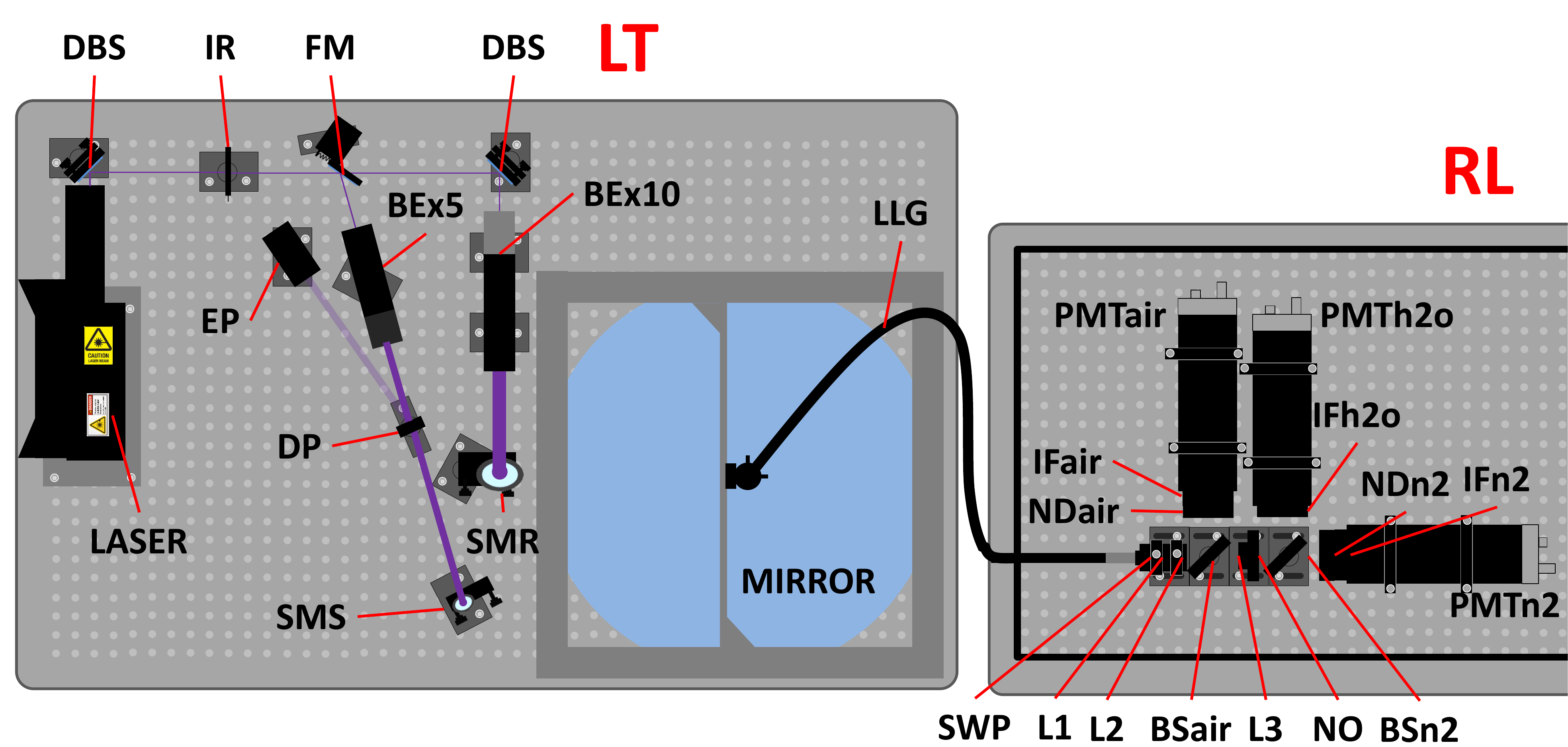}
    \caption{\label{fig:LT_RL}
      The laser transmitter and the Raman \lidar receiver layout (top view),
      ND\_air and ND\_\n2 are combinations of neutral density filters (see
      text).
    }
  \end{center}
\end{figure}

The optics of this system (Tab.~\ref{tab:1}) were arranged as shown in
Fig.~\ref{fig:LT_RL}. Dichroic beam splitting mirrors (DBS) removed the
residuals of the primary and secondary harmonics. A motorized flipper mirror
(FM) was raised or lowered to switch between two beam paths. The LT Raman path
(LTR) was used to direct pulses at 100\,Hz into the sky for RL measurements. The
LT side scatter path (LTS) included a depolarizer (DP) and a pick-off energy
monitor. This path was used to direct pulses vertically at 4\,Hz for
measurements by the AMT. The properties of the two beams delivered to the sky
are listed in Tab.~\ref{tab:1}. Although a single beam path could have been used
in theory, in practice the use of separate paths simplified the operation and
alignment procedures considerably without compromising the scientific
objectives. The LTR beam direction was fine tuned to match the direction of the
optical axis of the RL mirror (which pointed in the nominal vertical direction).
Independently, the direction of the other path was set to vertical relative to a
laser level alignment device that was also used in the alignment of the AMT.

\begin{table}[t]
  \begin{center}
    \caption{\label{tab:1}
    Technical specification of the Laser Transmitter subsystem.
    }
    \begin{tabular}{l l}
      \toprule
      Laser                                  & Big Sky Laser Centurion Nd:YAG\\
      Wavelength                             & 354.7\,nm  \\
      Line width                             & $\sim$\,1\,cm$^{-1}$\\
      Spectral Purity                        &  $>$ 99$\%$ \\
      Pulse duration                         & 7\,ns \\
      Output energy                          & 6\,mJ (nominal)\\
      Divergence LTR/LTS                     & 0.3\,mrad/0.6\,mrad\\
      Polarization LTR/LTS                   & linear/randomized\\
      Rep. Rate LTR/LTS                      & 100\,Hz/4\,Hz\\
      \midrule
      1\inch dichroic beam splitter (DBS)    & CVI BSR-35-1025 \\
      Flipper mirror (FM)                    & Newport 8892-K and CVI BSR-31-1025 \\
      5x beam expander (BEx5)                & CVI BXUV-10.0-5X-354.7 \\
      10x beam expander (BEx10)              & Thorlabs ELU-25-10X-351\\
      2\inch steering mirror (SMR)           & Newport 20QM20EN.35 \\
      1\inch steering mirror (SMS)           & CVI BSR-35-1025 \\
      Depolarizer (achromatic) (DP)          & Thorlabs DPU-25-A \\
      \bottomrule
    \end{tabular}
  \end{center}
\end{table}

\subsection{Raman \lidar}

The receiving telescope of the Raman \lidar consists of a 50\,cm diameter $f/3$
parabolic mirror pointing vertically beneath a UV transmitting silica window and
a motorized roof hatch. A liquid light guide couples the light reflected from
the mirror to a three-channel receiver. Dichroic beam splitters direct this
light onto three photomultiplier tubes (PMTs) that are located behind
narrow-band optical filters. These isolate the three scattered wavelengths of
interest, 354.7\,nm (Rayleigh/Mie or elastic scattering), 386.7\,nm (Raman \n2
backscattering), and 407.5\,nm (Raman \h2o backscattering). The data acquisition
system uses analog (up to 80\,MHz A/D) and photon counting (maximum count rate
250\,MHz) acquisition modules. The LT and RL are located in a temperature
conditioned building, rain and wind sensors conditionally control the opening
and closing of the roof hatch, see Sec.~\ref{sec:slowcontrol}.

The RL receiver collects the fraction of the laser beam photons that are
backscattered by the elastic Rayleigh and Mie scattering processes, as well as
those photons that are inelastically backscattered by \n2 and \h2o Raman
scattering processes. A scheme of the receiver can be found in
Fig.~\ref{fig:LT_RL}. It consists of a parabolic mirror, a liquid light guide
(LLG) that transports the collected light into the detector box containing a
combination of dichroic beam splitters (BS), interference filters (IF), neutral
density filters (ND), a notch filter (NO), three field lenses (L1, L2, L3) that
are combined to collimate the light beam, and the detectors (PMTs). The optical
and spectral characteristics of the different components can be found in
Tab.~\ref{tab:2} and Tab.~\ref{tab:3}, respectively. The read out of the PMTs is
carried out by electronic acquisition cards based on FPGA technology. These
low-consumption and low-cost cards simultaneously record the signals in current
mode (A/D) and photon counting mode (PhC). Electronic specifications are listed
in Tab.~\ref{tab:4}. Details of the RL performance are discussed in the next
three subsections.

\begin{table}[t]
  \begin{center}
    \caption{\label{tab:2}
      Technical specification of the Raman \lidar receiver optics.
    }
    \begin{tabular}{ll}
      \toprule
      Telescope         & Marcon parabolic mirror \\
      Diameter          & 504\,mm \\
      Focal length      & 1500\,mm \\
      Coating           & MgF$_2$ and Al protection \\
      Surface quality   & $\lambda /8$ RMS \\
      \midrule
      Liquid light guide (LLG)   & Newport 77629 \\
      \midrule
      Light collimator (L1, L2, L3)          & Thorlabs lenses LA1951/LA1131/LA1986 \\
      Effective focal length at 354.7\,nm    & $\simeq$\,165\,mm \\
      \bottomrule
    \end{tabular}
  \end{center}
\end{table}

\begin{table}[t]
  \begin{center}
    \caption{\label{tab:3}
      Technical specification of the spectral features of the Raman \lidar receiver.
    }
    \begin{tabular}{ll}
      \toprule
      Air beam splitter (BS\_air)                  & Barr BS-R345-361nm \\
      Reflectance at 354.7\,nm                     & $\geq$\,99\% \\
      \midrule
      \n2 beam splitter (BS\_\n2 )                 & Barr BS-R407-T320-395nm \\
      Reflectance at 407.5\,nm                     & $\geq$\,99\% \\
      \midrule
      Air interference filter (IF\_air)            & Barr IF-CWL354.7-BW6nm \\
      Central wavelength and bandwidth             & 354.7\,nm and 6\,nm\\
      Transmittance at 354.7\,nm                   & $\geq$\,83\% \\
      \midrule
      \n2 interference filter (IF\_\n2 )           & Barr IF-CWL3867-BW10A \\
      Central wavelength and bandwidth             & 386.7\,nm and 1\,nm \\
      Transmittance at 386.7\,nm                   & $\geq$\,77\% \\
      \midrule
      \h2o interference filter (IF\_\h2o )         & Barr IF-CWL4075-BW10A \\
      Central wavelength and bandwidth             & 407.5\,nm and 1\,nm\\
      Transmittance at 407.5\,nm                   & $\geq$\,65\% \\
      \midrule
      Notch filter (NO)                            & Barr LWP-T-378/415nm \\
      Rejection at 354.7\,nm                       & Optical density $\geq$\,6 \\
      \midrule
      Short wavelength pass filter (SWP)           & Newport 10-SWP-500 \\
      \bottomrule
    \end{tabular}
  \end{center}
\end{table}

\begin{table}[t]
  \begin{center}
    \caption{\label{tab:4}
      Technical specification of the Raman \lidar receiver detectors.
    }
    \begin{tabular}{ll}
      \toprule
      2\inch Photomultipliers (PMT) & Electron Tubes 9828B \\
      Typical quantum efficiency & 35\% \\
      Gain & 2.5\,$\times$\,10$^6$ \\
      Single electron rise time & 2\,ns \\
      \midrule
      DAQ cards & Embedded Devices APC-80250DSP \\
      Channels & 2, analog (A/D) and photon counting (PhC) \\
      Single photon maximum count rate & 250\,MHz \\
      PhC bin temporal resolution & between 100 and 1000\,ns \\
      PhC dead time between channels & $<$\,1\,ns \\
      A/D acquisition & up to 80\,MHz \\
      A/D bandwidth & 20\,MHz\\
      A/D resolution & 12\,bit\\
      \bottomrule
    \end{tabular}
  \end{center}
\end{table}

\subsubsection{Optical Efficiency}

The LLG input end is centered on the parabolic mirror axis, in a position close
to the focal plane. In this configuration the telescope has a field of view
(half angle) of about 2.7\,mrad. The LLG output connects to the detector box,
where the transported light is directed by a combination of lenses onto the
dichroic beam splitters and the optical filters that separate the different
\lidar returns according to the wavelength. The RL receiver has been simulated
with the \zemax optical design program, assuming that the pointing direction of
the laser beam is perfectly aligned with the telescope axis. The design
parameters considered in the simulation are:

\begin{itemize}
  \item the laser divergence (0.3\,mrad half angle);
  \item the distance between the laser beam and the telescope axes (about
    310\,mm);
  \item obscuration effects of the telescope frame and of the LLG holder;
  \item an estimation of the refractive indexes of the LLG core and cladding
    materials;
  \item the length and core dimension of the LLG;
  \item the distance of the LLG input end above the infinity focus position of
    the parabolic mirror;
  \item the optical characteristics and the relative position of the lenses,
    dichroic beam splitters and optical (interference, neutral and notch)
    filters.
\end{itemize}

\begin{figure}[t]
  \begin{center}
    \includegraphics[width=12cm]{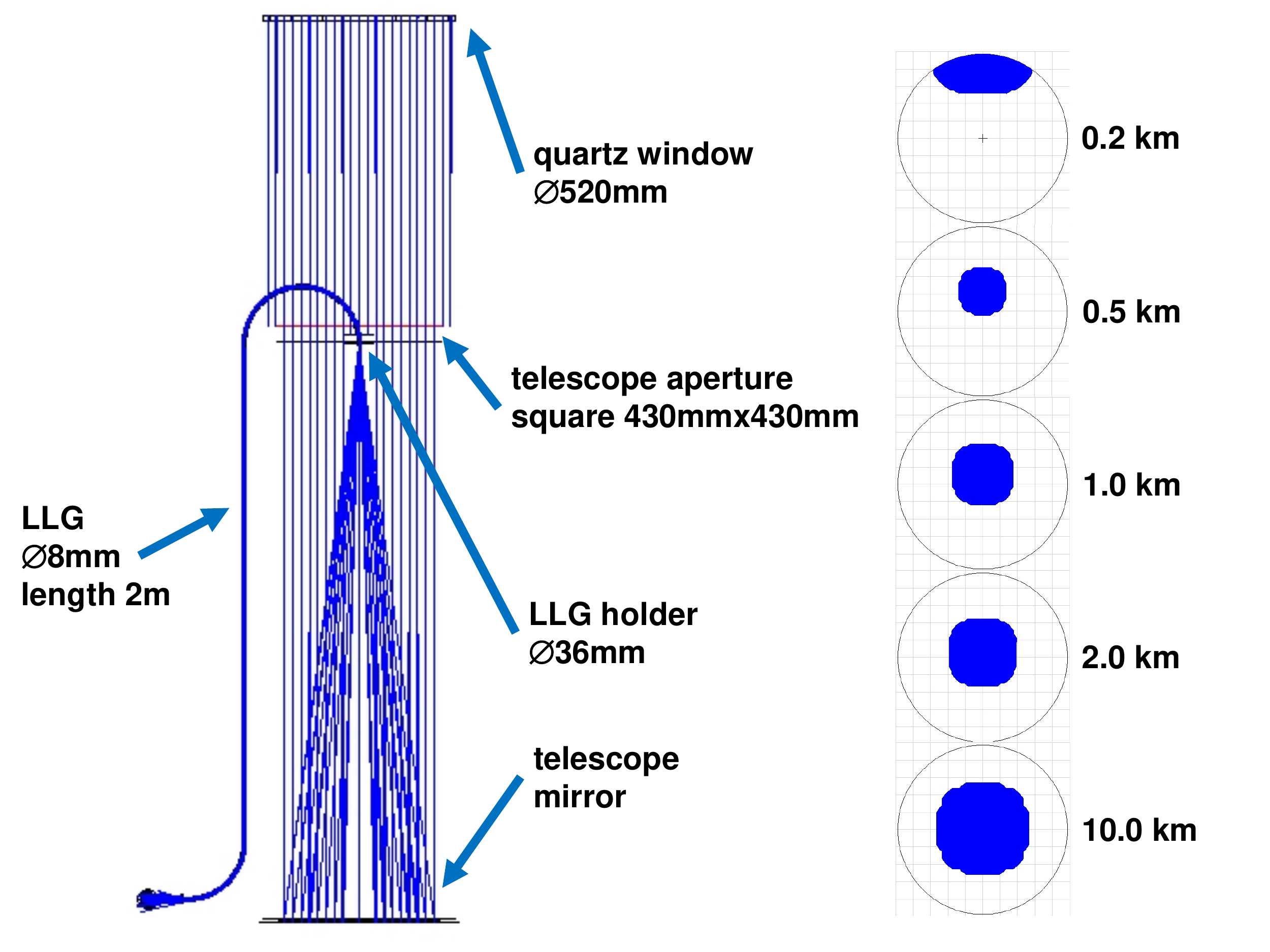}
    \caption{\label{fig:RL_zemax_1}
      Simulation of receiver telescope and LLG. Left: Layout of part of
      the simulation. Right: Laser beam image at the LLG input as a function of
      the distance from the telescope. The LLG input is positioned +9.0\,mm
      above the infinity focus position of the parabolic mirror.
    }
  \end{center}
\end{figure}

\begin{figure}[t]
  \begin{center}
    \includegraphics[width=12cm]{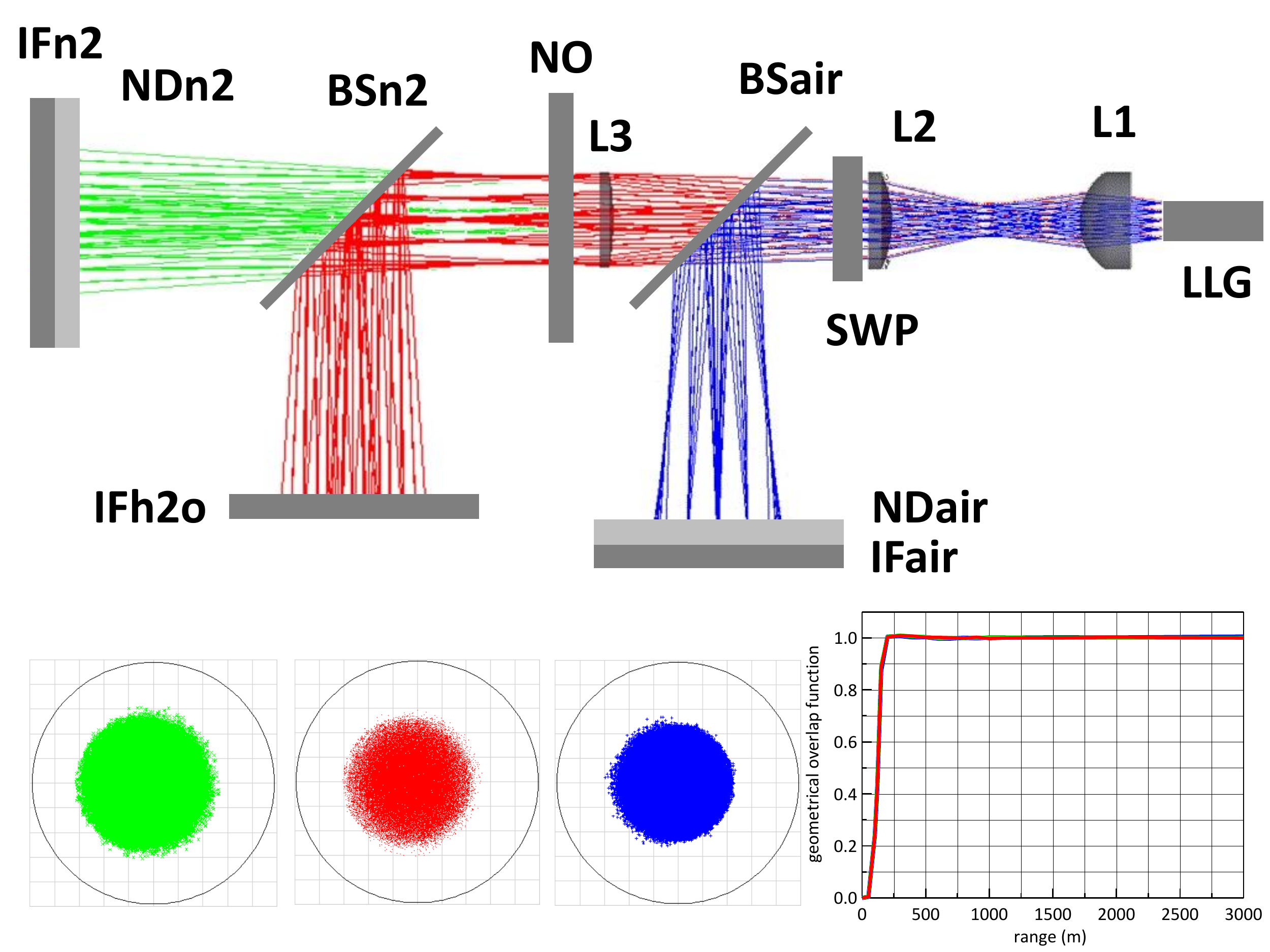}
    \caption{\label{fig:RL_zemax_23}
      Simulation of RL detector box. Top: Ray tracing simulation of the detector
      box. Bottom: Dimension of the light spots at wavelengths 354.7, 386.7 and
      407.5\,nm over the interference filters. In the lower right corner, the
      coincident geometrical overlap functions for the different RL detection
      channels are shown.
    }
  \end{center}
\end{figure}

The layout of part of the simulation is shown in the left panel of
Fig.~\ref{fig:RL_zemax_1}. The silica window, the telescope and LLG frames, and
the parabolic mirror, as well as the LLG can be represented in the simulation
with the real dimensions. The ray tracing calculation allows one to estimate the
optical efficiency of the system in collecting the \lidar returns from different
heights. The distance of the LLG input end above the infinity focus position of
the parabolic mirror is one of the parameters that can be optimized to improve
the optical efficiency of the RL receiver, i.\,e., the geometrical overlap
function. In the right panel of Fig.~\ref{fig:RL_zemax_1}, the laser beam image
at the LLG input as a function of the distance from the telescope is shown for
an LLG input position of +9.0\,mm above the infinity focus position of the
parabolic mirror. In this position, the geometrical collecting efficiency is
full (i.\,e., the entire image is transmitted into the LLG) from the lowest
range of about 300\,m. Upper and lower positions of the LLG end determine worse
overlap functions (i.\,e., full overlap at higher heights or a decrease in the
collection efficiency of light from higher altitudes). The laser beam image on
LLG has a maximum diameter $\leq$\,5.0\,mm and the impinging rays have an
incident direction that spreads over a numerical aperture (NA) of $\sim$\,0.16.
The propagation of the light (at 354.7, 386.7 and 407.5\,nm) along the LLG
preserves the NA. At the exit of the LLG, the light rays are distributed over
the entire LLG end diameter of 8\,mm, with exiting directions within an NA of
$\sim$\,0.15.

The ray tracing simulation of the detector box is illustrated by the optical
layout in upper part of Fig.~\ref{fig:RL_zemax_23}. In the lower part, the
dimension of the light spots at the different wavelengths over the interference
filters are shown. The 354.7\,nm light on the air interference filter (IF\_air)
falls within a circle of diameter $\sim$\,15.5\,mm, with an angle of incidence
that is $\in \left[0^{\circ}, 6^{\circ} \right]$. The \n2 Raman backscattered
photons on the \n2 interference filter (IF\_\n2 ) are diffused over a circle of
$\sim$\,17.0\,mm diameter, and the maximum angle of incidence is
$\sim$\,3.5\degree. For \h2o Raman backscattered photons the diameter of the
beam and maximum incident angle over the interference filter (IF\_\h2o ) are
$\sim$\,16.5\,mm and $\sim$\,3.5\degree, respectively. For such values of the
beam dimensions there are no important effects related to the spectral
characteristics of the dichroic beam splitters and interference filters.
Moreover, the light beams propagating into the RL detector box are slightly
diverging/converging by less than 6\degree off the normal incidence on the
interference filters. This causes only a small decrease of less than 0.1\% of
the central wavelength of the IFs. The decrease of the transmittance and the
increase of the bandwidth of the IFs are negligible~\cite{MacLeod}.

Finally, in the lower right corner of Fig.~\ref{fig:RL_zemax_23}, the relative
geometrical overlap functions for all \lidar channels as evaluated in the
simulation are shown. They are coincident, and the full overlap is achieved just
above a range of 250\,m.

\subsubsection{Spectral Features}

The atmospheric backscatter spectrum for an incident laser wavelength of
354.7\,nm is determined by the Rayleigh, Mie and Raman scattering processes. For
air at pressure $P$\,=\,101\,325\,Pa, temperature $T$\,=\,300\,K, a relative \n2
abundance of 0.781, and a content of water vapor of 10\,g\,kg$^{-1}$ of air, the
Rayleigh backscattered photons in general have a wavelength coincident with the
one of the incident photons (Cabannes line), but they also include photons
backscattered by pure rotational Raman scattering process of atmospheric
nitrogen, oxygen and other molecules. It has to be mentioned that in the
presence of atmospheric aerosols, their contribution (Mie scattering)
superimposes to the Rayleigh backscattered light and has a narrower spectral
width. About 98\% of Rayleigh backscattered photons are in the Cabannes line
that has a spectral width coincident with that of the laser line. The other
photons have wavelengths that cover a band of few nanometers ($\pm$\,3\,nm)
centered around 354.7\,nm. The Stokes vibration-rotation Raman backscattering of
the \n2 and \h2o molecules produce photons that have wavelengths around
386.7\,nm and 407.5\,nm, respectively. For \n2 Raman backscattering,
$\sim$\,86\% of the energy is at 386.7\,nm (Q branch), and the rest is in the O
and S side branches that span about 5\,nm around the central line; the complete
\h2o Raman backscattering is in 407.5\,$\pm$\,3\,nm~\cite{Wandinger, Chiao-Yao}.

\begin{figure}[t]
  \begin{center}
    \includegraphics[width=12cm]{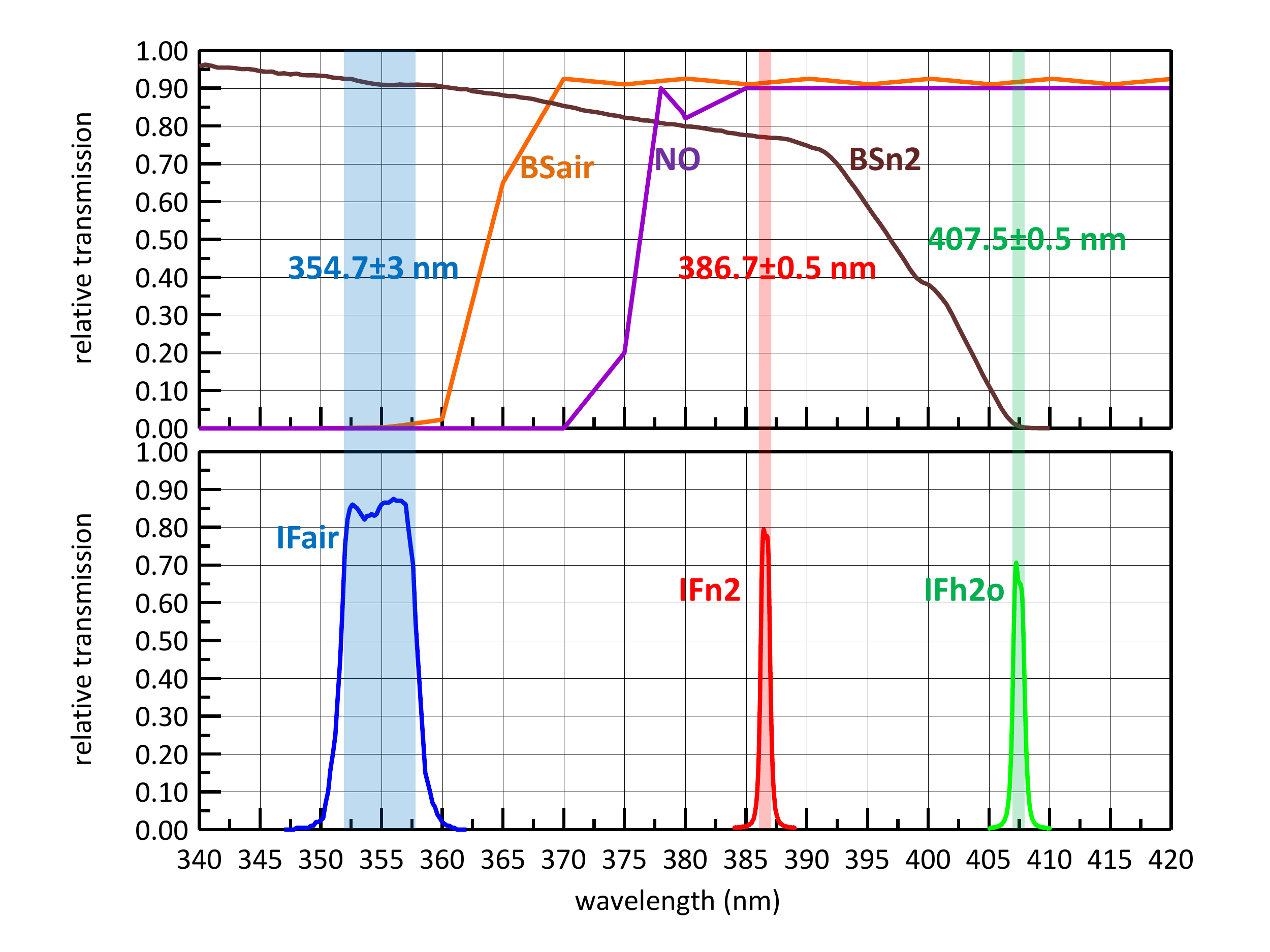}
    \caption{\label{fig:spectr}
      The spectral characteristics of the dichroic beam splitters (BS), the notch
      filter (NO), and of the interference filters (IF). The blue, red and green
      bands indicate the FWHH of the various IFs.
    }
  \end{center}
\end{figure}

The relative spectral transmissivities of the dichroic beam splitters, notch and
interference filters combined in the RL detector box are shown in
Fig.~\ref{fig:spectr}. The blue, red and green bands represent the FWHH
(6.0\,nm, 1.0\,nm and 1.2\,nm for IF\_air, IF\_\n2 and IF\_\h2o ) of the
transmission curves of the interference filters (bottom panel). The BSs separate
the different wavelengths in a quite efficient way (i.\,e., BS\_air reflects
$\sim$\,99.8\% of 354.7\,nm and transmits $\sim$\,91.4\% and $\sim$\,91.7\% of
386.7 and 407.5\,nm photons). The NO prevents the propagation of elastic photons
in the Raman detection channels (the NO optical rejection ratio at 354.7\,nm is
$\sim$\,1:10$^6$, and the transmissivity at 386.7 and 407.5\,nm is
$\sim$\,90.0\%). The FWHH of the interference filters mainly determine how much
of the Rayleigh/Mie, \n2 and \h2o Raman backscatter spectrum is collected by the
corresponding \lidar channels: CH\_air, CH\_\n2 and CH\_\h2o . The convolution
of the relative spectral transmissions of the IFs with the Rayleigh/Mie and
Raman bands gives:
\begin{itemize}
  \item CH\_air  collects $\sim$\,99\% of the total Rayleigh/Mie backscatter spectrum,
  \item CH\_\n2  collects $\sim$\,95\% of the total \n2 Raman backscatter spectrum,
  \item CH\_\h2o collects $\sim$\,99\% of the total \h2o Raman backscatter spectrum.
\end{itemize}

The individual line strengths in the Raman spectra are temperature
dependent~\cite{Whiteman}. In principle, the collected \lidar backscatter
returns may be temperature sensitive, especially when only a portion of them is
detected. In our case, the percentage changes in the detected energy,
backscattered by the atmospheric medium with temperatures from 200 to 300\,K
(the typical range of temperature in the troposphere) is less than a few tenths
of a percent for the Rayleigh and Raman bands: our system has a detection
capability that is temperature insensitive.

The ratios between the Raman backscatter coefficients of \n2 and \h2o molecules
and the Rayleigh backscatter coefficient are $\sim$\,7.1\,$\times$\,10$^{-4}$
and $\sim$\,3.4\,$\times$\,10$^{-5}$, respectively. The spectral discrimination
of the different RL detection channels has to be quite efficient to avoid
cross-talk effects (i.\,e., the propagation of elastically backscattered photons
in the CH\_\n2 and CH\_\h2o ). The combination of the dichroic beam splitters
with the notch and interference filters (see Fig.~\ref{fig:spectr}) determines
the total spectral collection efficiency $\eta_{ij}$ of the different \lidar
returns ($i$\,=\,354.7, 386.7 and 407.5\,nm) in the RL channels
($j$\,=\,CH\_air, CH\_\n2 and CH\_\h2o ). $\eta_{ij}$ includes also the spectral
reflectivity of the telescope, the transmissivities of light guide, lenses, beam
splitters, interference filters, etc.. The values of $\eta_{ij}$ for the three
channels and wavelengths are listed in Tab.~\ref{tab:5}.

\begin{table}[t]
  \begin{center}
    \caption{\label{tab:5}
      $\eta_{ij}$ for the three receiver channels at each of the three wavelengths.
    }
    \begin{tabular}{r c c c}
      \toprule
      $\eta_{ij}$ & CH\_air & CH\_\n2 & CH\_\h2o \\
      \midrule
      354.7\,nm & $\sim 3.7\times 10^{-6}$  & $\sim 4.0 \times 10^{-21}$ & $\sim 1.3 \times 10^{-21}$ \\
      386.7\,nm & $\sim 1.3\times 10^{-12}$ & $\sim 1.3 \times 10^{-2}$  & $\sim 1.5 \times 10^{-6}$  \\
      407.5\,nm & $\sim 1.6\times 10^{-12}$ & $\sim 1.4 \times 10^{-8}$  & $\sim 3.7 \times 10^{-2}$  \\
      \bottomrule
    \end{tabular}
  \end{center}
\end{table}

The Rayleigh and Raman backscatter coefficients $\beta_{i}$ ($i$\,=\,354.7,
386.7 and 407.5\,nm) of air at a pressure of 101\,325\,Pa, a temperature of
300\,K, with a relative \n2 abundance of 0.781, and a content of water vapor of
10\,g\,kg$^{-1}$ are:
\begin{itemize}
  \item Rayleigh backscatter coefficient at 354.7\,nm $\simeq$
    8.2\,$\times$\,10$^{-6}$\,m$^{-1}$\,sr$^{-1}$,
  \item \n2 Raman backscatter coefficient at 386.7\,nm $\simeq$
    5.8\,$\times$\,10$^{-9}$\,m$^{-1}$\,sr$^{-1}$,
  \item \h2o Raman backscatter coefficient at 407.5\,nm $\simeq$
    2.8\,$\times$\,10$^{-10}$\,m$^{-1}$\,sr$^{-1}$.
\end{itemize}

The product $\eta_{ij}\beta_{i}$ is a measure of the signal intensity of the
backscatter $\beta_{i}$ in the RL channel $j$. The relative intensities of the
expected signals in CH\_air, CH\_\n2 and CH\_\h2o are $\sim 0.4$, $\sim 1.0$ and
$\sim 0.1$, respectively. In Tab.~\ref{tab:6}, $\eta_{ij}\beta_{i} /
\eta_{i(j=i)}\beta_{i}$ values are listed.

\begin{table}[t]
  \begin{center}
    \caption{\label{tab:6}
      Ratio of signal intensity of the backscatter for all combinations of the
      three receiver channels and the three wavelengths.
    }
    \begin{tabular}{r c c c}
      \toprule
      $\eta_{ij}\beta_{i} / \eta_{i(j=i)}\beta_{i}$ & CH\_air & CH\_\n2 & CH\_\h2o \\
      \midrule
      354.7\,nm & 1                         & $\sim 3.2 \times 10^{-19}$ & $\sim 3.7 \times 10^{-20}$ \\
      386.7\,nm & $\sim 3.6 \times 10^{-7}$ & 1                          & $\sim 4.0 \times 10^{-5}$  \\
      407.5\,nm & $\sim 4.3 \times 10^{-7}$ & $\sim 1.1 \times 10^{-6}$  & 1                          \\
      \bottomrule
    \end{tabular}
  \end{center}
\end{table}

The rejection of the ``out of band'' backscattering signals is quite good in
each RL channel. In CH\_air it is at least $2:10^8$, in CH\_\n2 it is better
with $1:10^6$. In CH\_\h2o , the rejection ratio at 354.7\,nm is $4:10^{20}$,
the \n2 Raman backscattered return can be up to 2\,ppm of the \h2o signal which
is negligible. In the testing phases of the RL receiver, we realized that
CH\_\h2o was affected by an unexpected problem. When the light at 354.7\,nm
propagates into the LLG, the materials that constitute the guide show a
fluorescent re-emission of light that reaches CH\_\h2o , and superimposes to the
\h2o Raman backscattering. The LLG consists of a plastic tube covered by a
protective aluminum spiral and covered by a PVC jacket. The inner tube is filled
with a proprietary, transparent, anaerobic, non-toxic fluid. The 8\,mm core is
sealed at both ends with polished fused silica windows and protected by an
interlocking stainless steel sheathing. The refractive indices of the inner
fluid and of the plastic tube are $\approx$\,1.42 and $\approx$\,1.35 at
354.7\,nm wavelength. An LLG similar to the one installed in the RL has been
tested in laboratory. It was discovered that the LLG, when illuminated with very
low intensity laser light from an excimer laser (at $\sim$\,351\,nm), shows a
fluorescence emission that appears like a line centered at $\sim$\,400\,nm with
a FWHH of $\sim$\,7\,nm. This measurement has been done with a commercial
spectrometer with a wavelength resolution of 1.5\,nm. It is likely that in our
RL the 354.7\,nm light could induce a fluorescence at $\sim$\,404\,$\pm$\,4\,nm
(shift of $\sim$\,3474\,$\pm$\,375\,cm$^{-1}$). These photons originate in the
LLG and can be easily detected in CH\_\h2o . This fluorescence is
instantaneously re-emitted. It can have a lifetime up to $\sim$\,100\,ns and it
has a spectral shift that is quite similar to the one that happens in aqueous
mixtures. This is in agreement with the probable composition of the LLG liquid
core of CaCl/H$_2$O, CsBr/CsI/H$_2$O or DMSO (dimethylsulfoxide) water
mixtures~\cite{patent1, patent2, patent3}. In summary, the signal detected in
CH\_\h2o contains a contribution coming from the LLG fluorescence.

For these reasons, we plan to replace the LLG in future measurements with a
fiber bundle (high OH content silica/silica) showing negligible fluorescence,
and we do not analyze the CH\_\h2o signal.

\subsubsection{\lidar Signal Sampling}

The backscattered photons in the RL receiver are confined in a collimated beam
that impinge on a central, relatively small area of the photocathodes of the
PMTs. This way it is likely that there are no effects due to the variation of
sensitivity with the position of incident light on the photocathodes. Each
detected photon individually generates a single photoelectron (PE). The typical
PEs are negative pulses with a mean amplitude of $\sim$\,75\,mV, and a typical
rise time of $\sim$\,3\,ns. The PMT outputs (a collection of PEs) can be
considered as a waveform superimposed over the shot noise of the photons or as a
random sequence of pulses originating from individual photons. The former leads
to analog recording (A/D), the latter to photon counting (PhC).

Our electronic acquisition card is based on FPGA technology and uses a fast
digital signal processor unit for both analog and photon counting detection. In
A/D mode, the PMT signal is digitized into an 8\,bit waveform at an adjustable
sampling rate. The duration of the single sample can be 12.5, 25, 50 or 100\,ns
and the waveform is reconstructed for a total of 1024~samples. This corresponds
to the detection of \lidar returns that extend to heights of 1.92, 3.84, 7.68 and
15.36\,km with a spatial resolution of 1.875, 3.75, 7.5 and 15\,m. In a standard
Raman \lidar run the A/D sample rate is set at 10\,MHz (100\,ns sample
duration). The first 10~samples (of 1024) are collected before the start of the
laser shots, they can be used to measure the signal background.

In PhC mode the PMT pulses are counted using a discriminator, its adjustable
threshold level allows to reject noise pulses. The best threshold level for the
output of our PMTs is $-$25\,mV. The formed pulses are counted in 1024
consecutive time bins, each bin can range in widths from 25 to 1000\,ns in 25\,ns
increments. This allows to collect the return signal along the time scale of
(1024~bins) $\times$ (time bin width). For the RL configuration, the bin width
is set at 200\,ns. This corresponds to a range resolution of 30\,m spanning up
to $\sim$\,30\,km height. The DAQ provides the sum of the signals integrated
over a certain number of laser shots. These data are saved on the DAQ system
memory board. An external computer is used to control the DAQ via USB
connection. A single data file for each of the RL channels cumulates about
12\,000 laser shots (at a laser repetition rate of 100\,Hz this takes
2~minutes). The data is stored as ASCII files. Each file contains information
on the system settings and the raw data in digit units as photon counts per bin
vs.\ time or averaged current waveforms vs.\ time. The PhC mode is preferable in
signal acquisition, but in the low range regions, close to the instrument, where
the PhC rate is higher than 10\,MHz, the A/D detection can be used to avoid
pile-up effects that can affect the PhC.

\begin{figure}[t]
  \begin{center}
    \includegraphics[trim= 2cm 0cm 2cm 0cm, width=.95\textwidth]{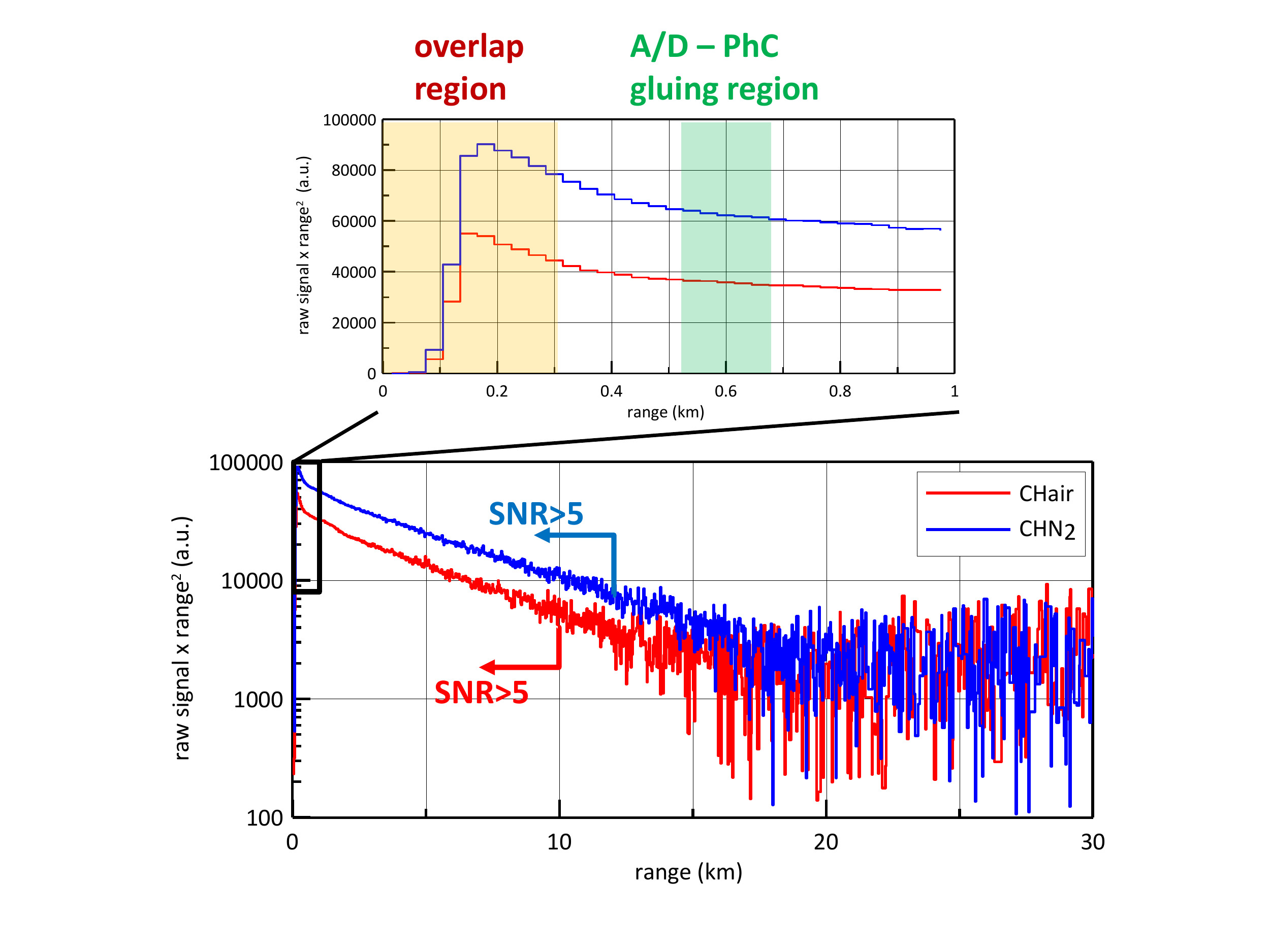}
    \caption{\label{fig:Lsign}
      The $L_{\rm E}(s)$ and $L_{\rm R}(s)$ signals in CH\_air and CH\_\n2 , a
      combination of the A/D and PhC detection are shown for a typical
      measurement session. The levels corresponding to SNR\,=\,5 are indicated.
      The areas where the geometrical overlap function has an effect and where
      the gluing between A/D and PhC is done are shown in the zoom of the range
      between ground and 1\,km height in the bottom panel.
    }
  \end{center}
\end{figure}

The $L_{\rm E}(s)$ and $L_{\rm R}(s)$ signals, accumulated for about 24~minutes
and representing one hour of common observation of RL and AMT are shown in
Fig.~\ref{fig:Lsign}. The ratio between $L_{\rm E}(s)$ and $L_{\rm R}(s)$ is
$\sim$\,0.5, comparable to the estimations of the expected signals in CH\_air
and CH\_\n2 . The gluing of the A/D and PhC detections has been done in the
range around 600\,m where the PhC rate is well below 10\,MHz for both $L_{\rm
E}(s)$ and $L_{\rm R}(s)$. The SNR falls below 5 above 10\,km for $L_{\rm E}(s)$
and above 12\,km for $L_{\rm R}(s)$. According to our design of the receiver we
expect that the geometrical overlap function modulates the \lidar returns from
ground up to 300\,m.

\subsection{Atmospheric Monitoring Telescope}

The AMT is a UV air fluorescence cosmic ray telescope that was optimized for
this project. The AMT was commissioned in January 2010. Regular measurements
started after a 10~month engineering phase on October~8, 2010. In total,
320~hours of data were recorded with the AMT over nine dark periods around new
moon. The decommissioning of the RL and the LT in July 2011 entailed the
termination of AMT measurements.

\subsubsection{Optics
\label{sec:AMToptics}}

A 3.5\,m$^2$ spherical mirror of four segments, PMT camera, and UV optical
filter comprise the AMT optics. They are the type used in the HiRes-II
detector~\cite{Boyer:HiRes}. As a light pulse from the laser travels upward
through the atmosphere, a reflected image of the scattered light reaching the
mirror sweeps downward across the PMT camera (Fig.~\ref{fig:AMTpic}, bottom
left). This light spot crosses the camera in about 30\,\us. The spot size is
about 2\,cm. The field of view of one PMT is 1\degree. A UV filter mounted on
the camera increases the contrast of UV light against the night sky background.

There is no direct line of site between the laser and the AMT. The Two Buttes
landmark, visible from both sites, was used as a survey reference to establish
the AMT azimuth direction. The AMT optical axis, defined as the direction of the
sky that is imaged on the center of the camera, was adjusted to 8.72\degree in
elevation and 16.0\degree in azimuth (CCW from North). After correcting for the
difference in altitude and the curvature of the earth, the AMT views the laser
between 1.5\,km and 10.6\,km above ground at the LT (2.7\,km and 11.8\,km
above sea level, respectively).

\subsubsection{Infrastructure
\label{sec:AMTinfrastructure}}

\begin{figure}[p]
  \begin{minipage}[t]{.671\textwidth}
    \centering
    \includegraphics*[width=.99\linewidth]{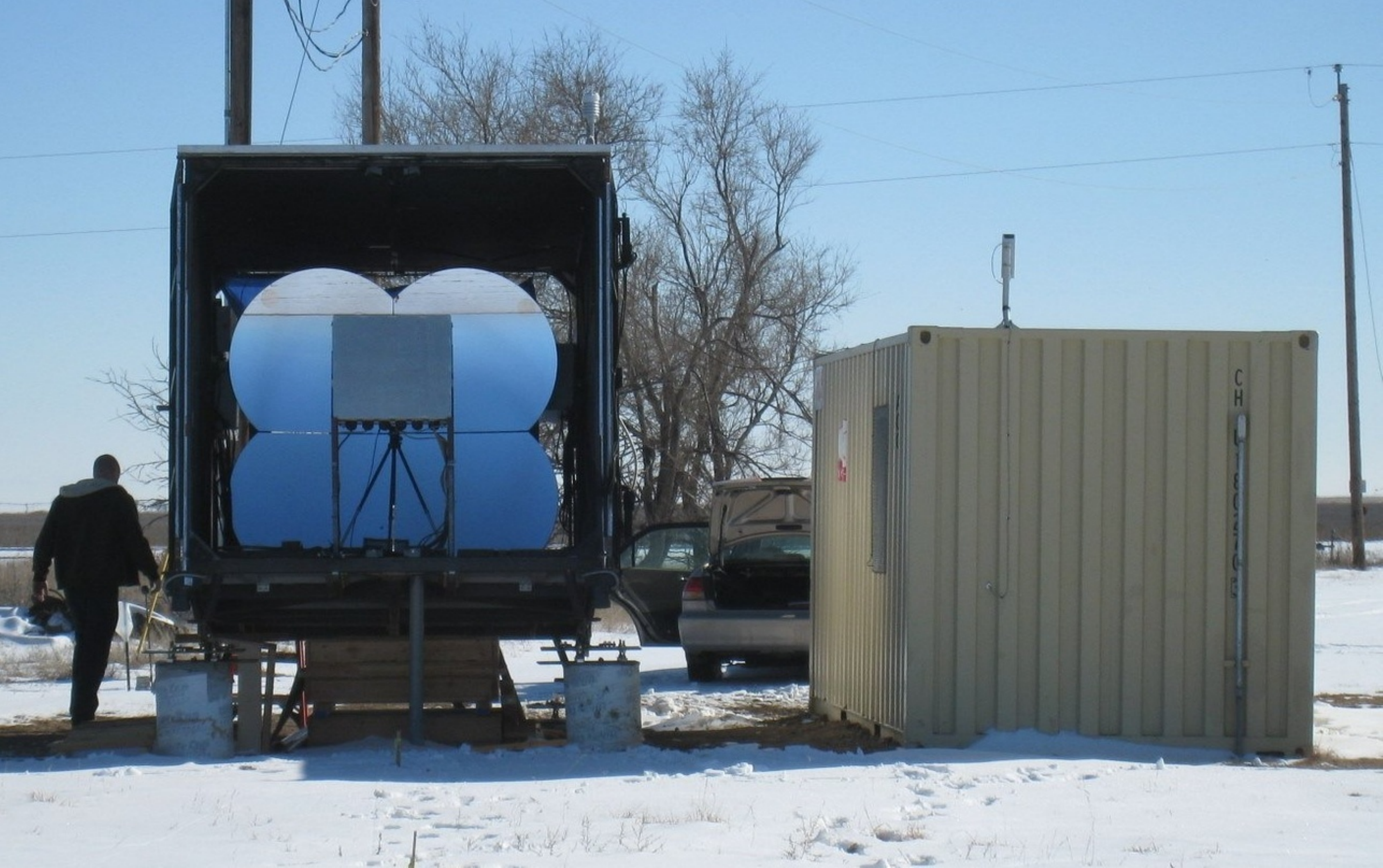}
  \end{minipage}
  \hfill
  \begin{minipage}[t]{.315\textwidth}
    \centering
    \includegraphics*[width=.99\linewidth]{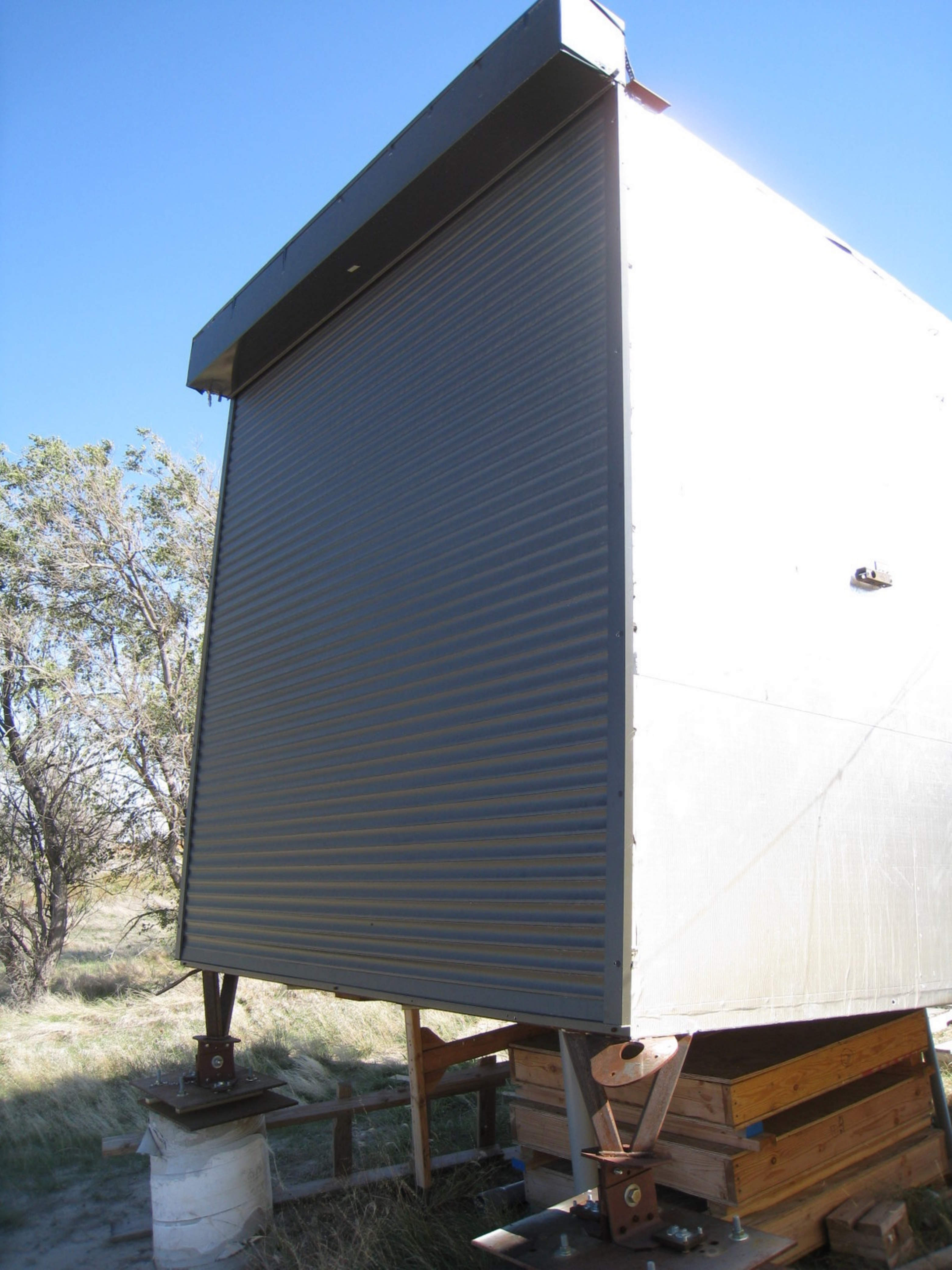}
  \end{minipage}
  \begin{minipage}[t]{.385\textwidth}
    \centering
    \includegraphics*[width=.99\linewidth]{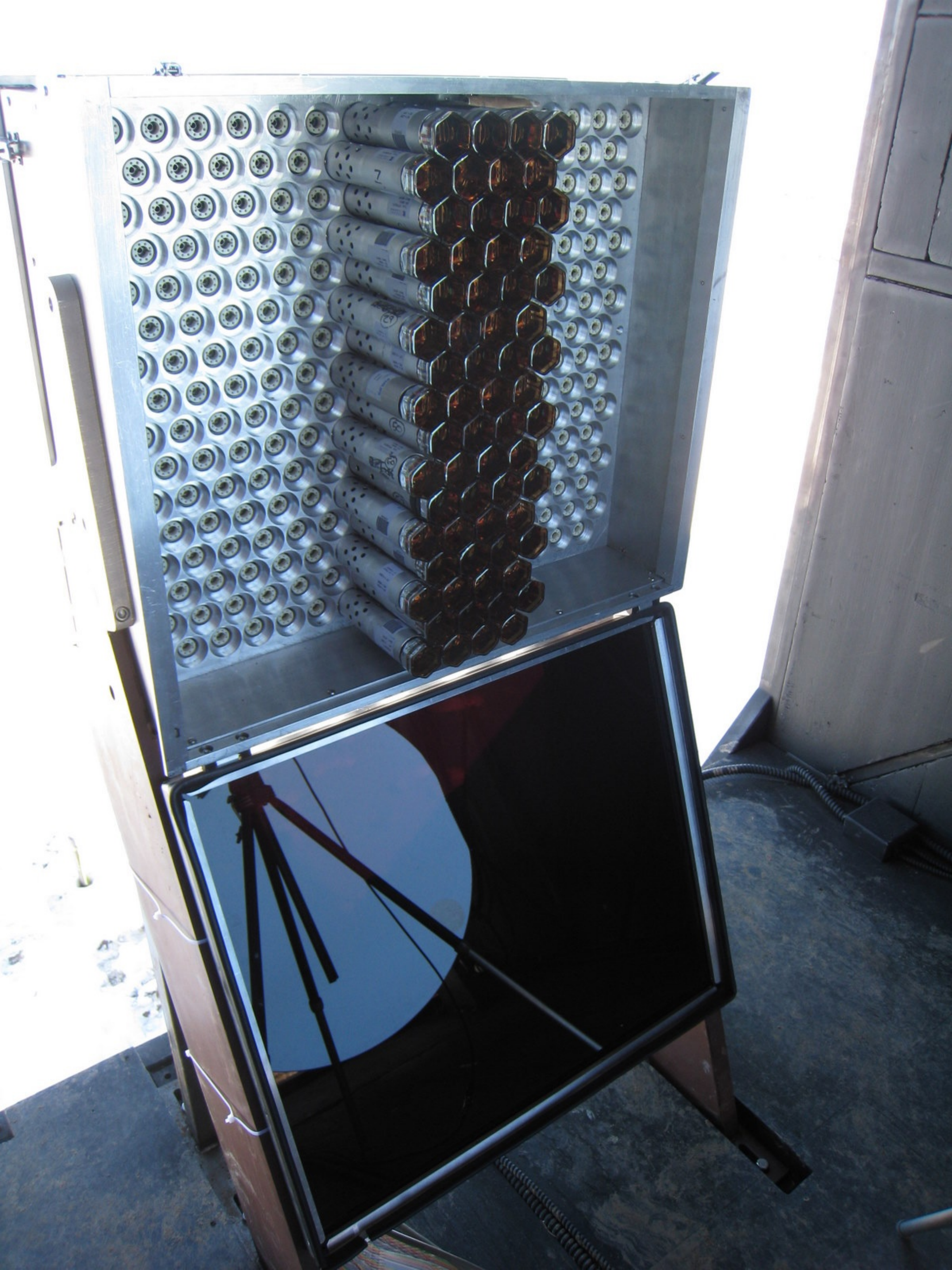}
  \end{minipage}
  \hfill
  \begin{minipage}[t]{.595\textwidth}
    \centering
    \includegraphics*[width=.99\linewidth]{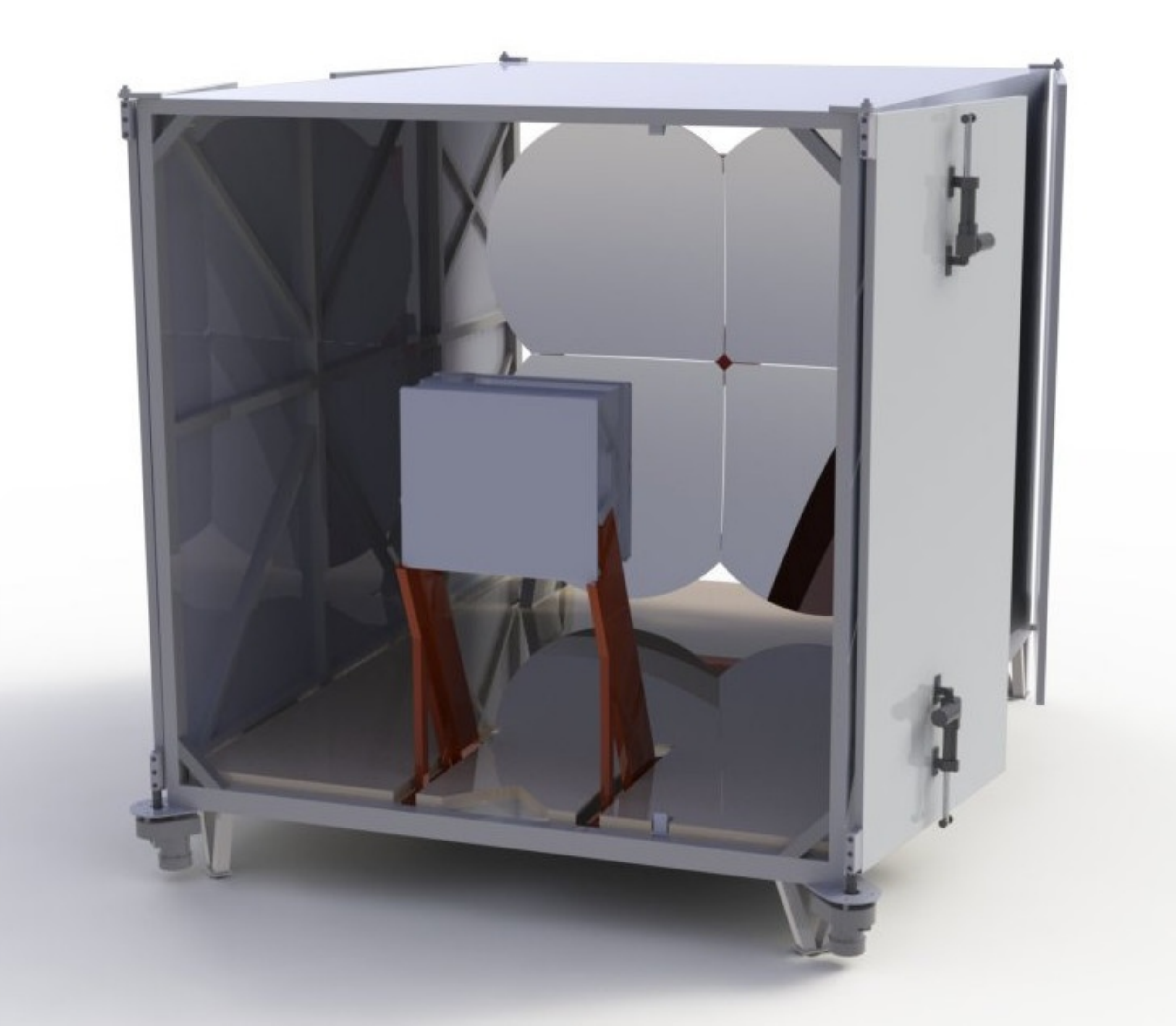}
  \end{minipage}
  \caption[]{\label{fig:AMTpic}
    AMT mirror, camera, weather sensors and calibration system are housed in a
    custom shelter (top left), the adjacent container contains the data
    acquisition system. A roll-up door can be closed across the entrance
    aperture (top right). The AMT camera (bottom left) is shown with the UV
    bandpass optical filter opened. A 3D rendering of the AMT using the CAD
    design software SolidWorks\textregistered{} (bottom right).
  }
  \vspace{10pt}
  \begin{minipage}[t]{.32\textwidth}
    \centering
    \includegraphics*[width=.99\linewidth]{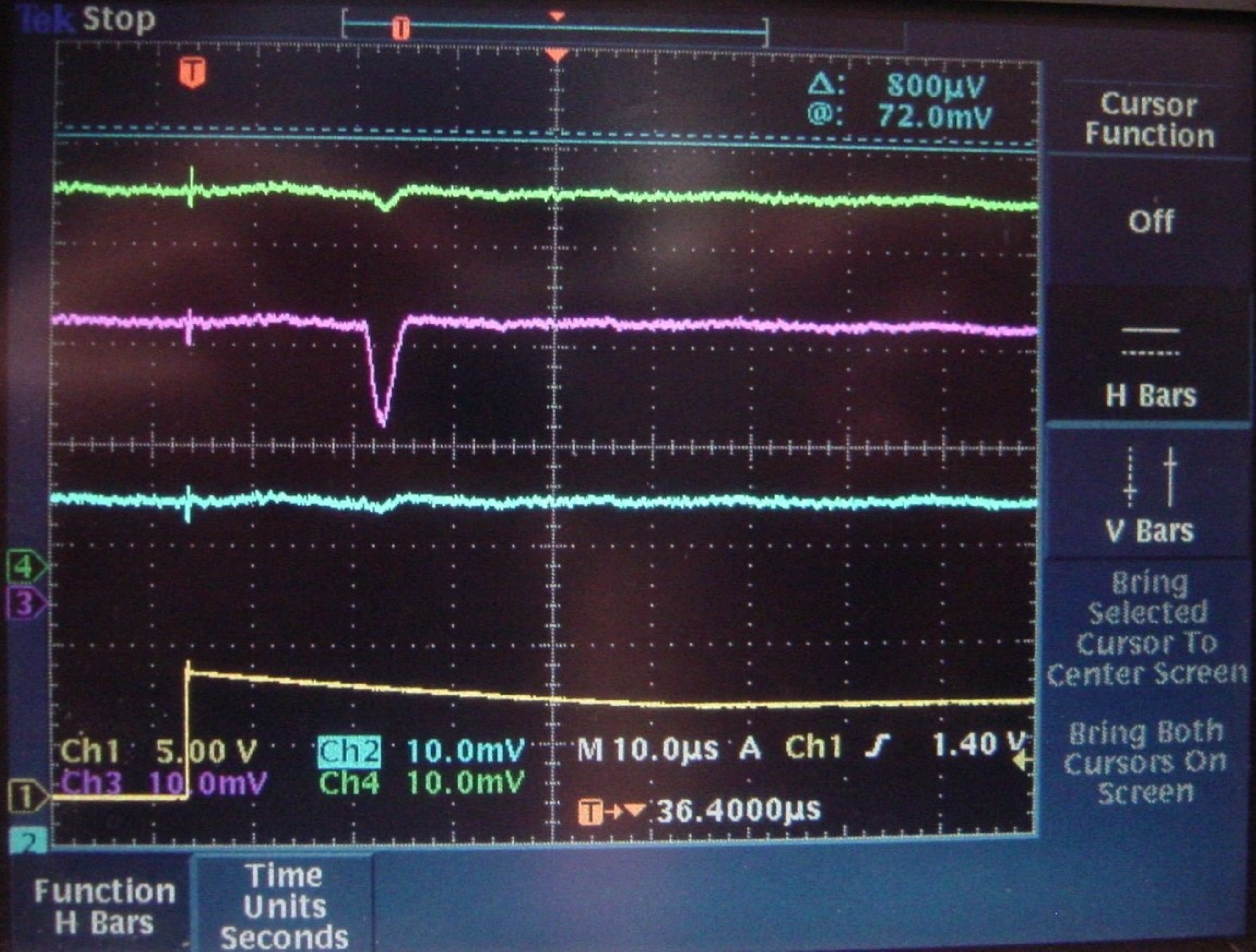}
  \end{minipage}
  \hfill
  \begin{minipage}[t]{.318\textwidth}
    \centering
    \includegraphics*[width=.99\linewidth]{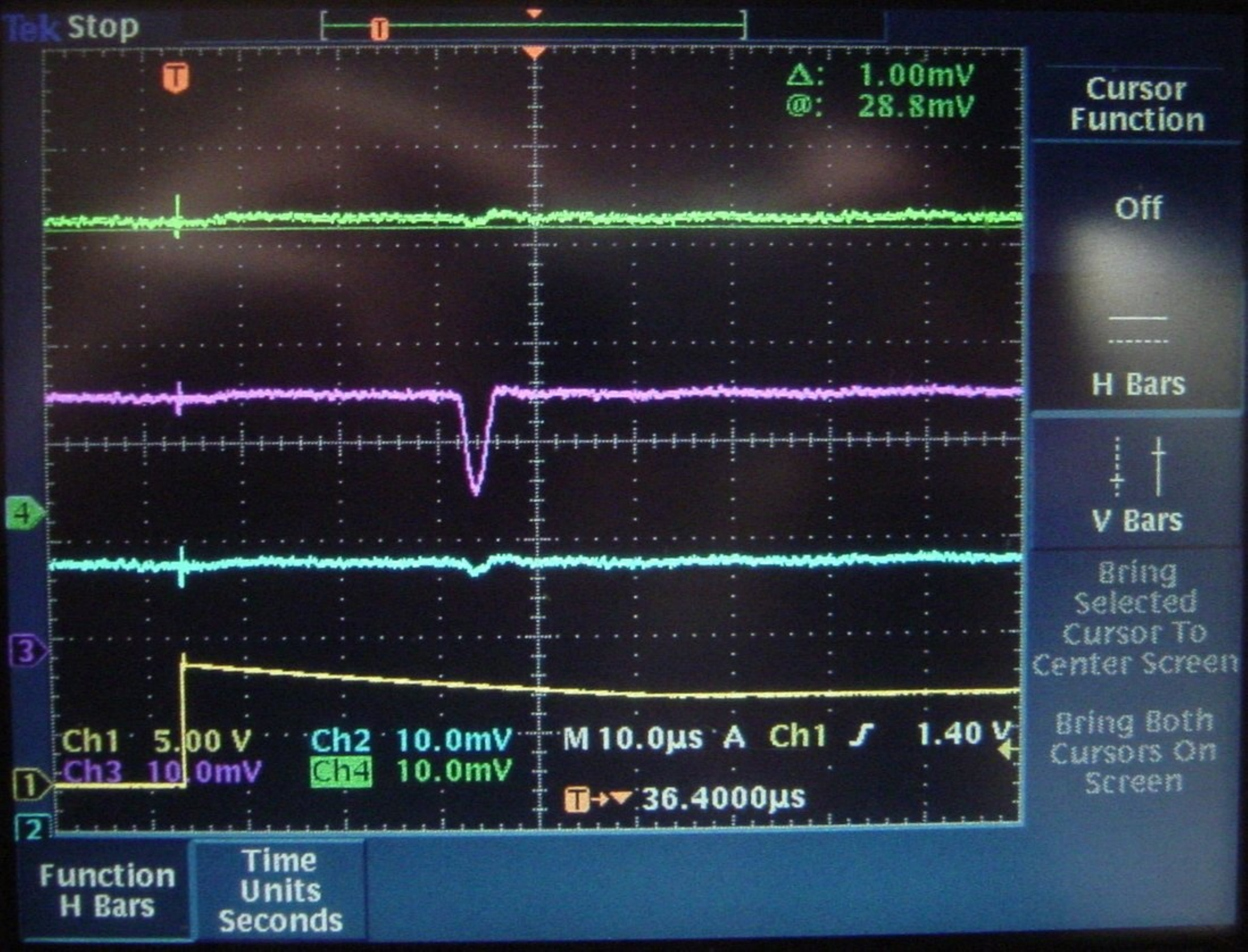}
  \end{minipage}
  \hfill
  \begin{minipage}[t]{.31\textwidth}
    \centering
    \includegraphics*[width=.99\linewidth]{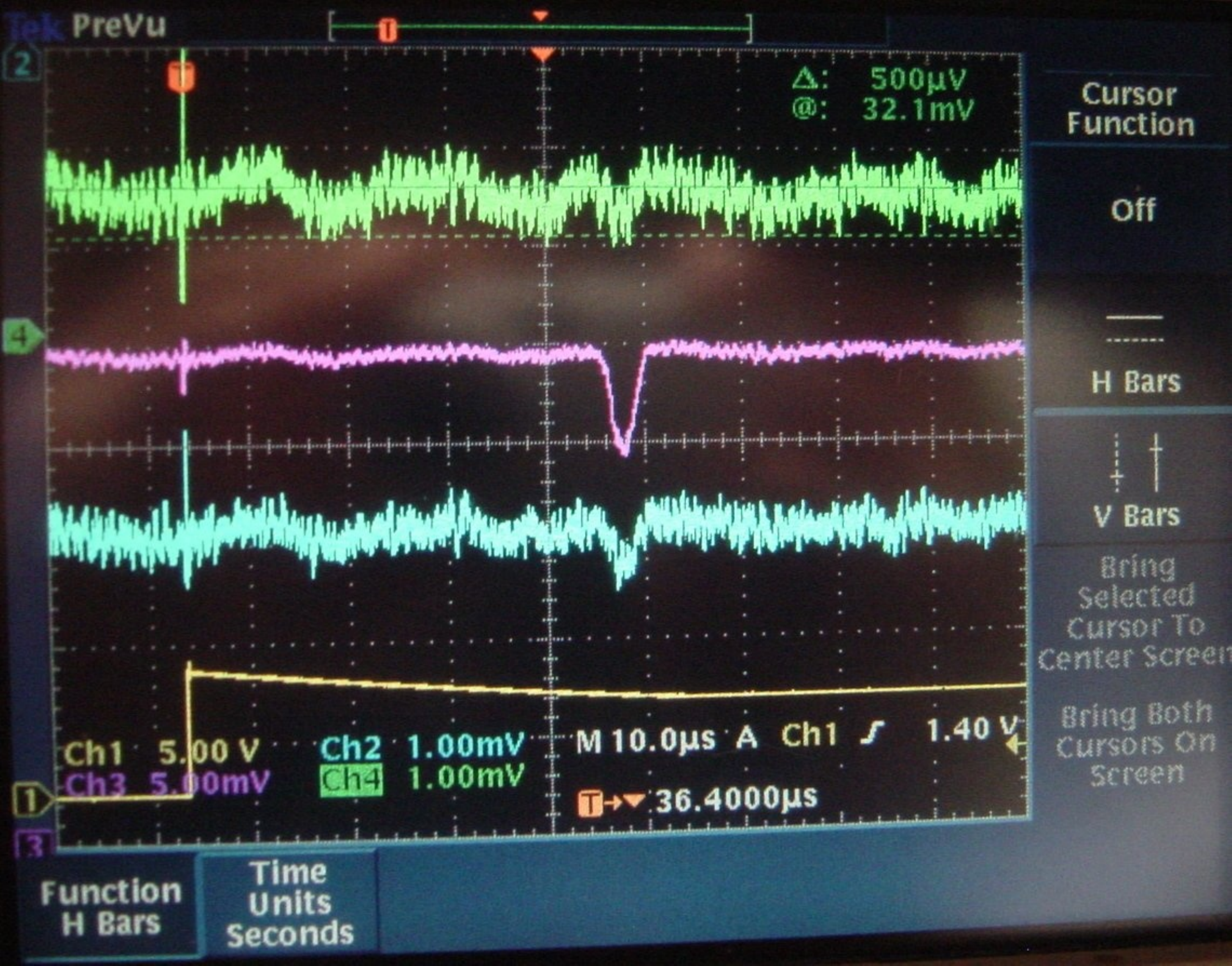}
  \end{minipage}
  \caption[]{\label{fig:firstDLFshot}
    First LT signals seen at the AMT. Left to right correspond to heights of
    2.0, 4.9 and 8.5\,km a.g.l. Each panel shows 3 signal traces, averaged over
    128 laser pulses, by PMTs in the same row. As the path length gets longer,
    signals shift further in time relative to the external trigger pulse (bottom
    trace).
  }
\end{figure}

The optical components are housed in a custom-built shelter. In
Fig.~\ref{fig:AMTpic}, a picture (top left panel) and 3D rendering (bottom right
panel) are shown. A roll-up door (top right panel) spans the end of the shelter
that forms the optical aperture. Since the shelter points north and away from
the sun, this door can be opened safely any hour of the day. A precipitation and
ultrasonic wind sensor are read by the slow control system to ensure the door is
closed during rainy or windy conditions. The shelter is mounted on four concrete
posts with fixtures that permit adjustments to the AMT pointing direction by a
few degrees. The AMT is aligned so that the vertical laser track passes near the
center of the optical field of view. Adjacent to the AMT shelter is a modified
shipping container. This ``counting house'' holds the data acquisition system
and portions of the slow control system. Power and signal cables connect the AMT
and the counting house through underground conduits. A local company provides a
wireless internet connection to the AMT from a broadcast tower about 4\,km away.
The AMT door and field of view can be observed remotely by the webcam.

\subsubsection{Camera and External Trigger}
\label{sec:camera}

\begin{table}[t]
  \begin{center}
    \caption{\label{tab:AMT}
      Technical specifications and spectral features of the PMTs and bandpass filter.
    }
    \begin{tabular}{ll}
      \toprule
      Photomultiplier model                        & Photonis XP3802 \\
      Photocathode type                            & Flat hexagonal bialkali \\
      Wavelength sensitivity of photocathode       & 300--650\,nm \\
      Quantum efficiency at 375\,nm                & 30\% \\
      Gain (operated)                              & $\sim 10^{5}$ \\
      \midrule
      UV bandpass filter transmission window       & 300--400\,nm \\
      Transmission at 354.7\,nm                      & $\sim$ 80\% \\
      \bottomrule
    \end{tabular}
  \end{center}
\end{table}

The AMT camera is designed to hold 256 PMTs arranged in a 16\,$\times$\,16
hexagonal grid. The camera, including the mechanical structure, the PMT/preamp
assemblies that are plugged into the camera backplane, and the LV and HV distribution,
were developed for the HiRes-II detector~\cite{Boyer:HiRes}. Some specifications
on the PMTs can be found in Tab.~\ref{tab:AMT}. For this project, the central 3
columns that viewed the LT vertical track were instrumented
(Fig.~\ref{fig:AMTpic}, bottom left panel) for a combined field of view of
3.5\degree in azimuth and 14\degree in elevation. These 48~PMTs were selected
from a larger set that was gain sorted. The photocathode was held at ground with
AC coupling (500\,\us time constant) applied to the anode signals. The output
voltage is the output current multiplied by 3\,k$\Omega$. The differential
preamp output was read out via equal length twisted pair cables.

A UV bandpass glass filter is mounted in front of the PMTs (some features in
Tab.~\ref{tab:AMT}). Tests at the Colorado site showed that this filter reduced
the night sky background light by about a factor of 2 while transmitting about
80\% of the 354.7\,nm laser light, which is in agreement with previous
measurements~\cite{Wiencke:1999}.

Oscilloscope traces of first light as recorded with individual PMT assemblies in
the AMT camera and averaged over 128 shots from the LT, can be seen in
Fig.~\ref{fig:firstDLFshot}. For atmospheric studies 200 laser shots were
averaged. The oscilloscope was placed in the counting house and measured the
pulses across the twisted pair cables. The first light test also verified the
external triggering scheme. A custom GPS module triggered the laser firing time.
A second, identical GPS module was placed at the AMT and programmed with a delay
to account for the light travel time from the laser to the AMT. This module
triggered the oscilloscope during the engineering phase and later the data
acquisition system.

Data from a temperature controlled UV LED system at the mirror center and from a
vertical nitrogen laser scanned across the field of view were used to flat field
the camera. Variable gain preamps in each channel of the front end of the data
acquisition electronics were adjusted to do this. The LED system follows the
design used in the \pao FD drum calibration system \cite{Brack:Cal}. During
routine nightly operation, the relative calibration of the PMTs was monitored
using the LED system.

\subsubsection{Data Acquisition
\label{sec:amtdaq}}

The camera readout is performed by the pulse shaping and digitization
electronics that are also used by the High Elevation Auger Telescope (HEAT)
extension~\cite{Mathes:2011icrc} at the \pao. The AMT readout includes one
Second Level Trigger (SLT) board and three First Level Trigger (FLT)
boards~\cite{Abraham:2009fd}. The readout is triggered externally. Each trigger
prompts the SLT to read out the FLTs and store the event. One FLT board with
22~channels is used per 16~pixel vertical column and records 100\,\us of data in
2000 50\,ns time bins. The readout software interface program is a modified
version of the run control program used for the FD telescopes of the \pao.

The DAQ timestamps AMT events so that the recoded traces can be matched to the
laser shots fired by the LT. The time is synchronized through the internet once
every night using a Network Time Protocol (NTP) server. To keep the time aligned
during the run, a one pulse per second signal (1~PPS) is provided by a GPS
module.

\begin{figure}[t]
  \begin{center}
    \includegraphics*[width=.99\linewidth,clip]{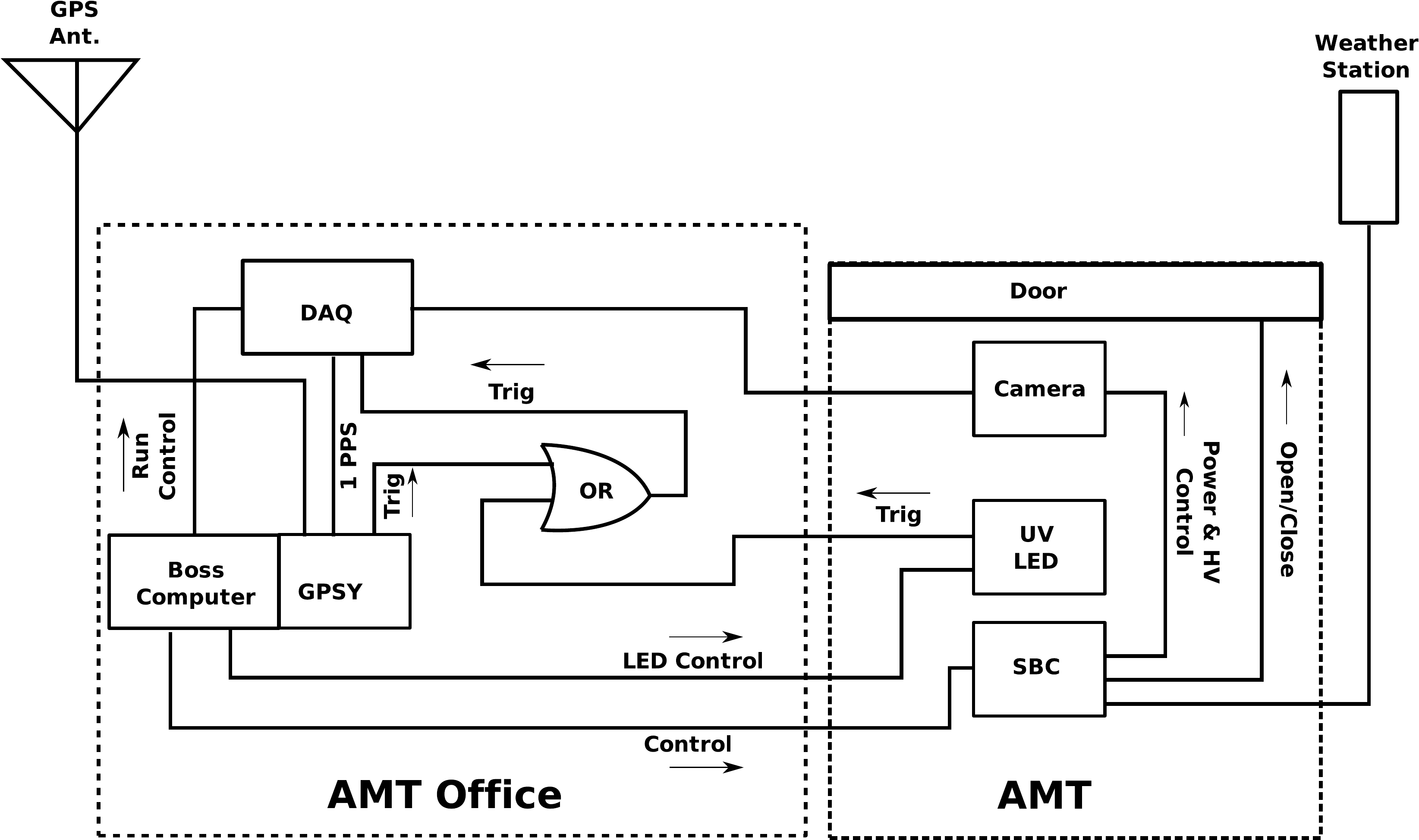}
    \caption[]{\label{fig:amtschem}
      A schematic of the AMT slow control system.
    }
  \end{center}
  \vspace{5pt}
\end{figure}

\subsubsection{Slow Control System
\label{sec:slowcontrol}}

Three Single Board Computers (SBCs) control the operation of the AMT. Their main
functions and the associated components are shown in Fig.~\ref{fig:amtschem}.
The SBCs and the DAQ are connected via an internal network. The two SBCs inside
the AMT shelter are responsible for the slow control systems and the LED
calibration system. The LED SBC can fire the LED at different pulse lengths and
widths. The system delivers a trigger signal for every LED pulse to trigger the
DAQ and records voltage traces for every pulse.

The AMT SBC controls and monitors the door, the low voltage power of the camera,
the high voltage of the PMTs and the weather station. The weather station
provides a set of temperature, pressure, wind, and precipitation every
5~seconds. A safety daemon program running in the background on the AMT SBC
checks if the values for wind and rain are within the safety margins. The SBC
shuts the door and powers down the high voltage if wind of more than
7\,m\,s$^{-1}$ or rain is measured.

In the counting house, the ``Boss'' SBC controls the nightly operations. From
there, commands can be sent to both the AMT and the LED SBC. A daemon on the
Boss reads a set of commands from a text file that specify the nightly
operations sequence. The Boss SBC includes the GPS module used to trigger the
AMT data acquisition.

The power supply of the Boss, the DAQ and all electronics inside the AMT are
secured by UPS, similar to the systems at the LT and RL. In case of main power
failure, enough power is available to close the AMT door and shut all systems
down securely. All systems are also connected to remote power control units that
can switch the power on or off and are controlled via the network by the Boss
SBC or manually by an operator.

\section{Data Collection and Analysis
\label{sec:operations}}

The AMT, RL, LT and various subsystems are all operated under computer control.
Their nightly operation is sequenced by automation scripts initiated on moonless
nights from the Colorado School of Mines campus. Operation and data collection
are then monitored remotely by collaborators in Colorado, Germany, and Italy.

\subsection{Sequence of Operations
\label{sec:seqop}}

\begin{figure}[t]
  \begin{center}
    \includegraphics*[width=.8\linewidth,clip]{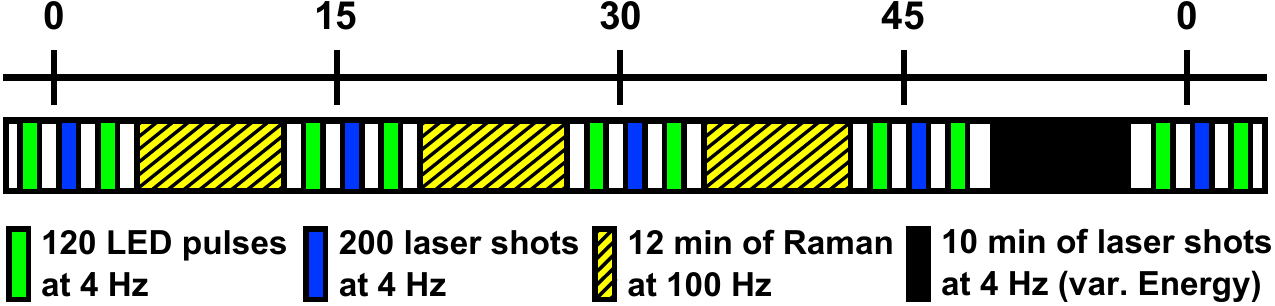}
    \caption[]{\label{fig:seqoperations}
      Scheme of the sequence of operations for one hour during combined data
      taking. 200~laser shots for the AMT are produced every 15\,minutes
      starting at minute~1 of every hour. 120~LED calibration pulses are done
      before and after every laser set. In the first three quarter hour breaks,
      the Raman \lidar collects data for about 8\,minutes. In the fourth quarter
      hour space, the laser is fired in AMT mode at different energies.
    }
  \end{center}
\end{figure}

The RL system and AMT have independent procedures which run in parallel, a
scheme of the hourly sequence is shown in Fig.~\ref{fig:seqoperations}. The
procedures are synchronized by GPS. Each night, the Raman \lidar system
operations start at 2:15~UTC and shut down at 11:00~UTC\footnote{Local time in
Colorado is UTC--7:00 in summer and UTC--6:00 in winter}. Starting at 3:00~UTC,
200 vertical shots for the AMT measurements are fired into the sky on minutes 1,
16, 31 and 46 of each hour. The laser fires at 4\,Hz, at 100, 350, 600, and
850\,ms after the full second. The data acquisition of the AMT is triggered
accordingly by the GPS board of the Boss computer, corrected for the time the
light takes to reach the AMT, around 130\,\us. In the Raman mode, the laser
fires at 100\,Hz. Backscattered photons from these laser shots are measured by
the Raman system, which collects data for eight minutes during the first three
intervals between sets of AMT laser shots. Between minutes 47 and 57, vertical
laser shots at lower energies are shot into the sky to check the linearity of
the response of the AMT camera.

The moon causes an intolerable level of interference for AMT data collection.
Therefore, the hours of AMT operation are dictated by the moon phase. For each
night, the hours in which the moonlight background is low are calculated and an
AMT command list is produced to perform operations during these hours. During
the hours of operation, 120~LED shots are fired before and after each set of AMT
laser shots. The LED sends its own trigger to the DAQ each time it fires. The
Raman system can operate every night, moonlight does not hinder the operations.

\subsection{Monitoring and Nightly Calibration}

For the relative calibration of the AMT camera during nightly operations, a UV
LED was mounted in the center of the mirror, pointing along the optical axis.
Inside this calibration light source, a temperature-controlled UV LED produces
short light pulses at 375\,nm. The LED was calibrated in the lab relative to
calibrated photodiodes by NIST. The LED is of the same type used to calibrate
the fluorescence detectors at the \pao~\cite{Abraham:2009pm}. Several layers of
diffuser material are used to make the light almost isotropic before it reaches
the camera. These pulses can be used to keep track of changes in the PMT
response. The LED is fired 120 times at a frequency of 4\,Hz, one minute before
and after the laser shots from the LT, providing two sets of calibration data
for each set of laser shots.

The current going through the LED is measured. Comparing the input signal with
the output generated by the PMTs as measured by the data acquisition system, a
relative calibration constant can be computed to correct the PMT signal. Before
calculating this constant, the PMT response has to be corrected for two
geometrical effects. The UV LED can be considered as a point source, the camera
body with the PMTs is flat, resulting in a smaller signal for PMTs further away
from the camera center. The flat camera body also causes a lowered effective
area A$_\mr{eff}$ of the outer PMTs. Accounting for both effects reduces the
observed light intensity of the LED to
\begin{linenomath*}
\begin{equation*}
  I_\mr{obs}(\theta) = I_0 \cdot \frac{A_\mr{eff}}{4\pi r^{\prime\,2}}
                     = I_0 \cdot \frac{A_\mr{PMT} \cdot \cos^3\theta}{4\pi r^2},
\end{equation*}
\end{linenomath*}
where $r$ is the distance between the LED and the center of the camera,
$r^\prime$ is the distance to a pixel offset from the center vertically by an
angle $\theta$.

\begin{figure}[htbp]
  \begin{center}
    \includegraphics*[width=1.3\linewidth,clip,angle=270]{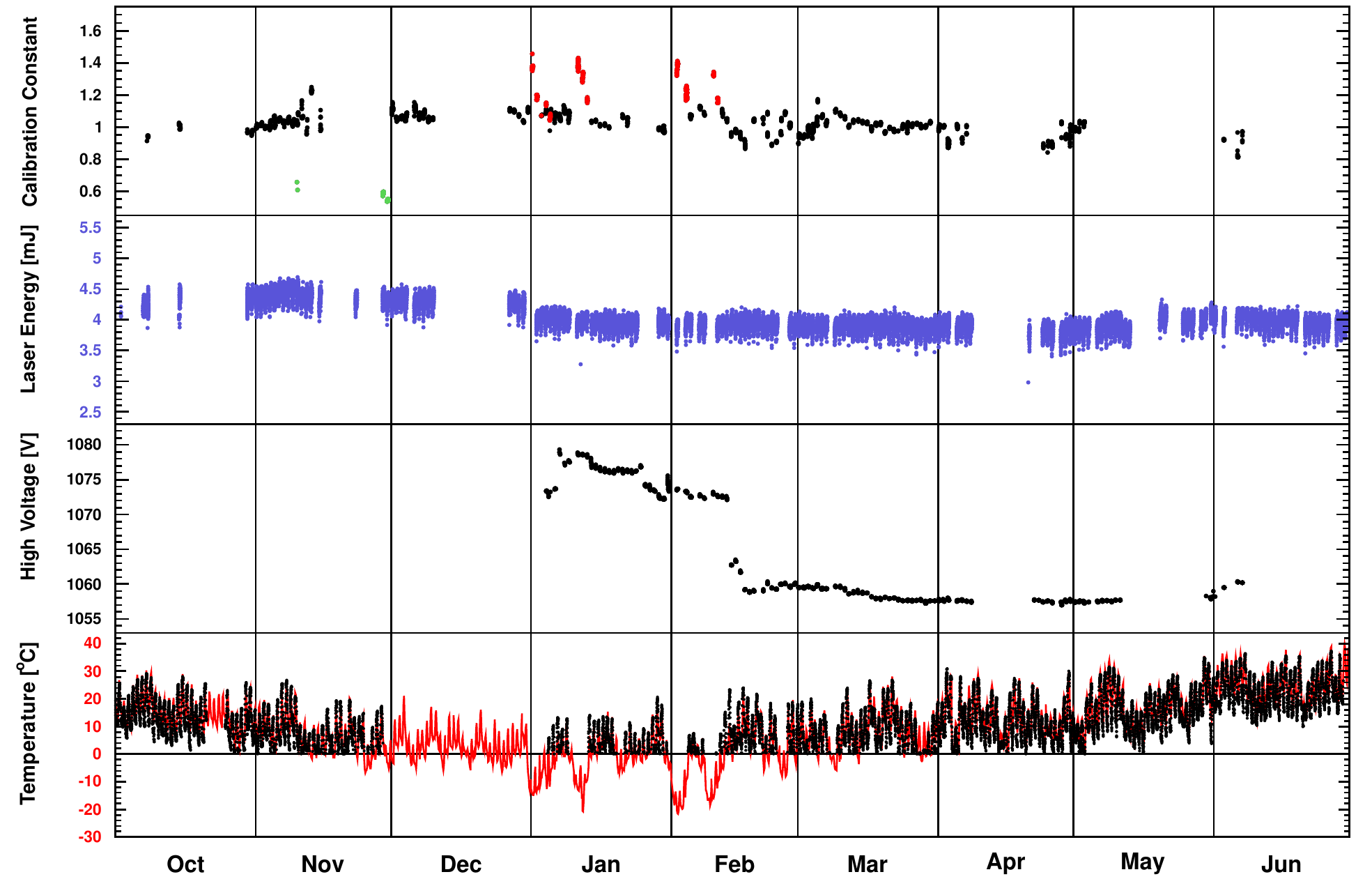}
    \caption[]{\label{fig:monitoring}
      Calibration and monitoring data for the AMT and the LT. In the top panel,
      the calibration constant for one of the pixels is shown. In the second and
      third panel, the laser energy and the readout from the high voltage
      monitor can be seen. On the bottom, the temperature of the weather station
      at the AMT is shown, together with meteorological model data in red. For
      more details see text.
    }
  \end{center}
\end{figure}

The currents of the UV calibration LED are averaged for every calibration set of
120~shots at 4\,Hz, lasting 30~seconds in total. In some rare cases, the
calibration light source did not fire or the DAQ was not triggered. If neither
the calibration before nor the one after a laser run was recorded, the closest
in time from a different set of laser shots is used, usually about 15~minutes
before or after. The actual calibration constant is calculated for each AMT
pixel by dividing the averaged voltage of the LED by the averaged PMT signal for
every 120 shot calibration set. This value is then normalized to the calibration
constant of an arbitrarily chosen night of data taking to get calibration
constants around unity. An absolute calibration of the pixels is not necessary,
the algorithm used to extract the aerosol optical depth \vaod from the measured
laser shots $SL$ relies on the comparison of the measurement to a reference
clear night $L_{\rm S}^{\rm clean}$, see Eq.~\eqref{eq:datanorm}. Therefore,
only the relative calibration of the current and the reference night have to be
known, any absolute calibration cancels out.

In Fig.~\ref{fig:monitoring}, some monitoring data are presented for the entire
duration of combined data taking. In the top panel, the calibration constant for
one of the AMT pixels in the central column with a field of view closest to the
horizon are shown. The average calibration constant is 1.04 with a standard
deviation of 0.09. In the second panel, the laser energy as measured with the
pick-off probe at the LT. The energy output was stable over the whole
operational period. In the third panel the output of a high voltage probe at the
AMT camera is displayed. It was installed in January 2010. In February of 2010 a
drop in HV is visible. There is no clear explanation for this change, but it is
expected that this drop does not impact the measurements.

Also shown in Fig.~\ref{fig:monitoring}, bottom panel, is the temperature as
measured with the weather station at the AMT (black dots). Since the available
data have several gaps when the station was not functioning or the temperature
was below zero, data from a global meteorological model
(GDAS~\cite{GDASinformation}) are used to supplement the data (red line). When
comparing the calibration constant with the temperature, it is obvious that
unusually high calibration constants correlate with particularly low
temperatures. This is most likely due to the calibration LED. Although it is
temperature controlled, it seems not to function well below a temperature of 
$-$8\degC. For this reason, data recorded under such conditions was
discarded, calibration constants for these nights are marked in red in the top
panel. In early and late November, wrong gain settings for the preamp of the
front end electronics were used accidentally, those calibration constants are
plotted in green. Periods of low temperatures and wrong gain settings are
discarded as bad periods.

\subsection{Collected Data}

\begin{figure}[t]
  \begin{center}
    \includegraphics*[width=.5\linewidth,clip]{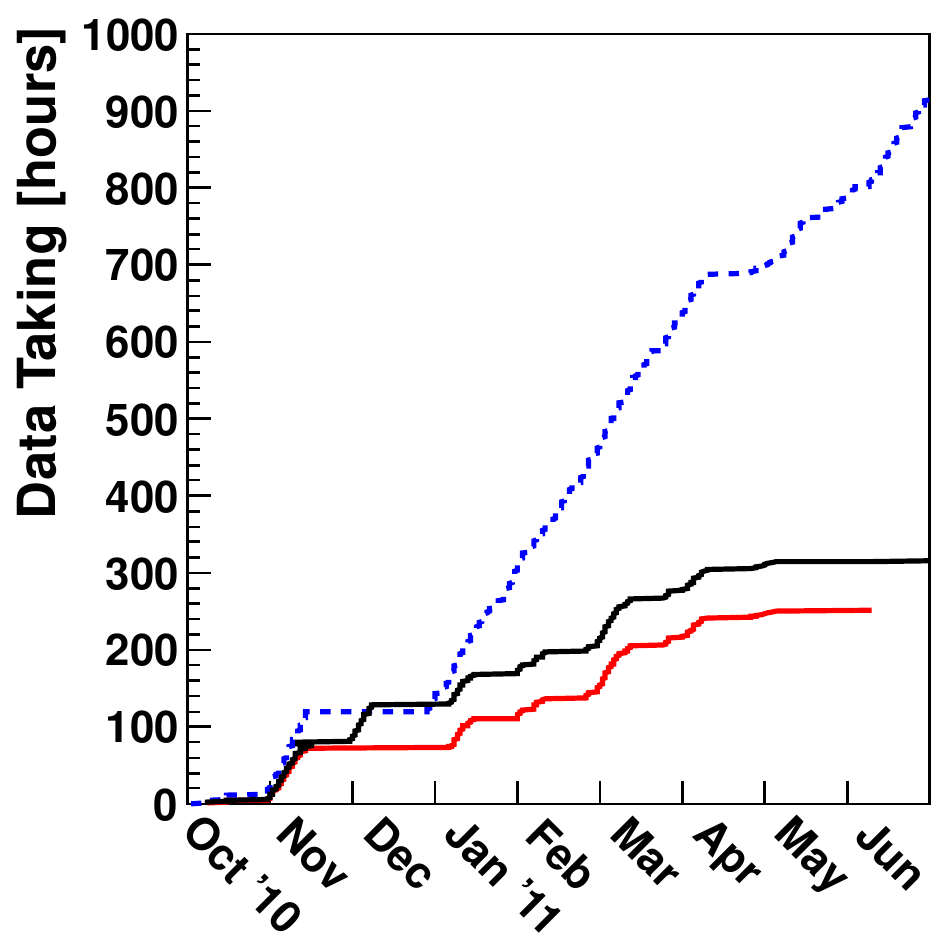}
    \caption[]{\label{fig:daqhours}
      Cumulative number of operational hours of the Raman \lidar (blue dashed)
      and the AMT (black). Hours where both systems acquired data are drawn in
      red.
    }
  \end{center}
\end{figure}

Since each system is capable of performing its startup, data collection and
shutdown procedures without an on-site shifter, these systems are monitored
remotely. Operations began on October~8, 2010, the systems have been monitored
from Golden, Colorado (approximately 500\,km from the site), from Karlsruhe,
Germany, from L'Aquila, Italy and once from \mal, Argentina. The longer period
without data taking in November 2010 was due to a failure of the Raman DAQ. The
AMT can only operate in nights with low illuminated moon fraction, so about
1.5~weeks before and after new moon. Regular operations were disturbed by broken
components like the weather station or the high voltage power supply. Towards
the end of measurements in June 2010, the operation of the AMT became more and
more difficult, in May 2011 data from two hours and in June 2011 only one hour
is available. It was decided to stop operations of the AMT one month earlier,
the Raman \lidar was decommissioned in July. In total, 320~hours of data were
collected by the AMT, 937~hours by the Raman alone and 251~hours of combined
data are available, see Fig.~\ref{fig:daqhours}. Rejecting bad periods due to
bad weather and other causes, 292 AMT and 233 combined hours remain.

The Raman \vaod and the backscatter coefficient $\beta_\mr{aer}$ were estimated
using the $L_{\rm E}(s)$ and $L_{\rm R}(s)$ signals in CH\_air and CH\_\n2 ,
that are a combination of the A/D and PhC detection along a Raman \lidar
measurement session. The hourly GDAS molecular atmosphere~\cite{GDASinformation}
corresponding to the measurement period is used to estimate the contribution of
the Rayleigh scattering processes into the \lidar returns. The GDAS data are
available in 3-hourly, global, 1\degree latitude-longitude (360\degree by
180\degree) datasets. Each data set consists of surface data and data for 23
constant pressure levels (from sea level up to about 26\,km). Among the
meteorological fields contained in the GDAS hourly file there are temperature,
pressure and relative humidity values from which the molecular number density
profile of the atmosphere can be estimated for calculating the Rayleigh
backscatter and extinction coefficients as used in equations
\eqref{eq:alpharaman}--\eqref{eq:betaraman}. The GDAS grid point closest to the
site of the Raman \lidar is 38\degree north and 102\degree west, which is East
of the town of Lamar.

For this particular analysis, no optical overlap correction was used, both \vaod
and $\beta_\mr{aer}$ profiles are valid in a range between about 0.5\,km and
5--6\,km above ground level. Raman \lidar data were collected during clear and
cloudy periods. In total, from 937~hours of data taking, after the quality
check, 930 reconstructed hourly profiles of \vaod and simultaneous
$\beta_\mr{aer}$ are available, among those profiles about 280 present low
clouds.

\begin{figure}[htbp]
\begin{center}
  \begin{minipage}[t]{.99\textwidth}
    \centering
    \includegraphics*[width=.99\linewidth]{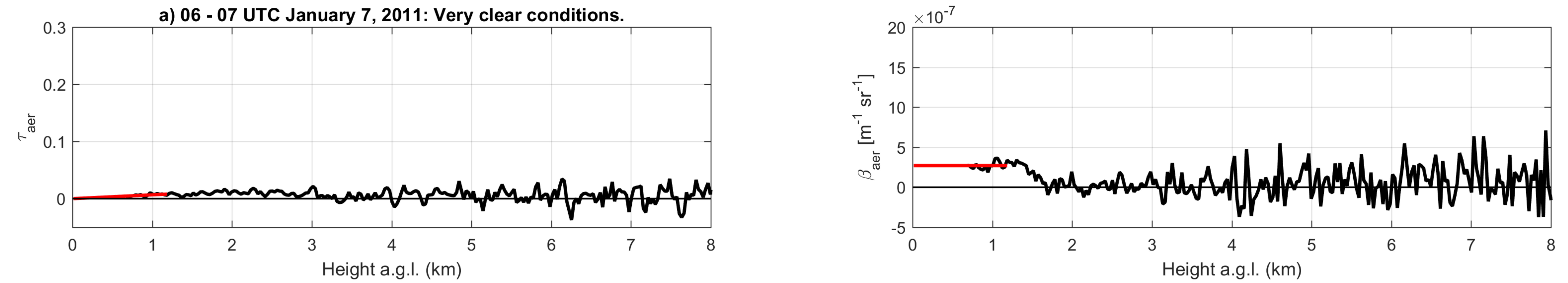}
  \end{minipage}
  \begin{minipage}[t]{.99\textwidth}
    \centering
    \includegraphics*[width=.99\linewidth]{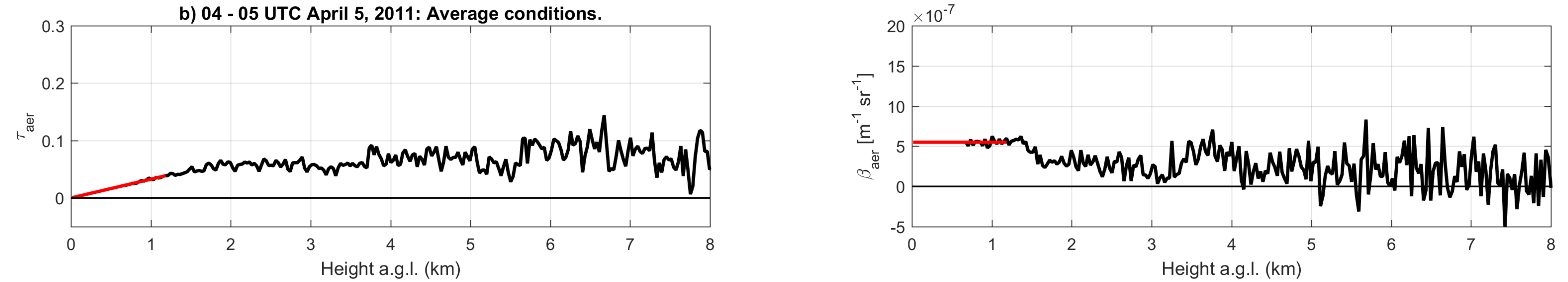}
  \end{minipage}
  \begin{minipage}[t]{.99\textwidth}
    \centering
    \includegraphics*[width=.99\linewidth]{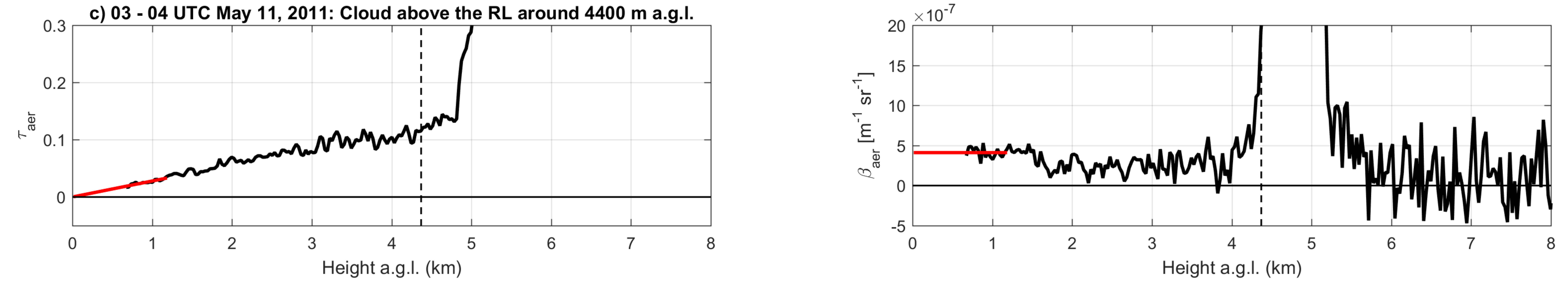}
  \end{minipage}
  \caption[]{\label{fig:ramprofiles}
    The raw data from the Raman \lidar analysis shown in black, in red the
    extrapolation to ground. On the left, the vertical aerosol optical depth
    \vaod, on the right, the aerosol backscatter coefficient $\beta_\mr{aer}$.
    From top to bottom, the Raman data measured during extremely clear (a),
    average (b) and cloudy conditions (c) are shown.
  }
\end{center}
\end{figure}

In Fig.~\ref{fig:ramprofiles}, as examples, the raw profiles are drawn in black.
The profiles of \vaod and $\beta_\mr{aer}$ are measured during very clear (a)
and average (b) conditions, and, in the bottom panel (c), in presence of clouds,
as it can be seen in $\beta_\mr{aer}$. A crude cloud height determination was
introduced, if $\beta_\mr{aer}$ rises above
2\,$\times$\,10$^{-6}$\,m$^{-1}$\,sr$^{-1}$, the minimum cloud height is set,
visualized in the panel (c) by a black dashed line.

Every hour during AMT operations, four sets of 200 laser shots are fired by the
laser. A set is fired every 15~minutes, resulting in 800 shots per hour. The
signal of every pixel is corrected by the calibration constant, then the traces
of all pixels are summed and an hourly average is formed.

To analyze the measured data, a reference clear night had to be identified in
order to calculate the aerosol optical depth. All available profiles were
checked for the highest photon number at the aperture. Two hours in the same
night were identified, the chosen reference profiles were measured on January~7,
2011 between the hours of 6 and 8~UTC. This choice was verified by the Raman
\lidar data of this night which also shows a quite low aerosol load,
see top panel in Fig.~\ref{fig:ramprofiles}. Two hours are sufficient for
defining a reference night, the number of shots included is 1600.

\begin{figure}[htbp]
  \begin{center}
    \begin{minipage}[t]{.99\textwidth}
      \centering
      \includegraphics*[width=.99\linewidth]{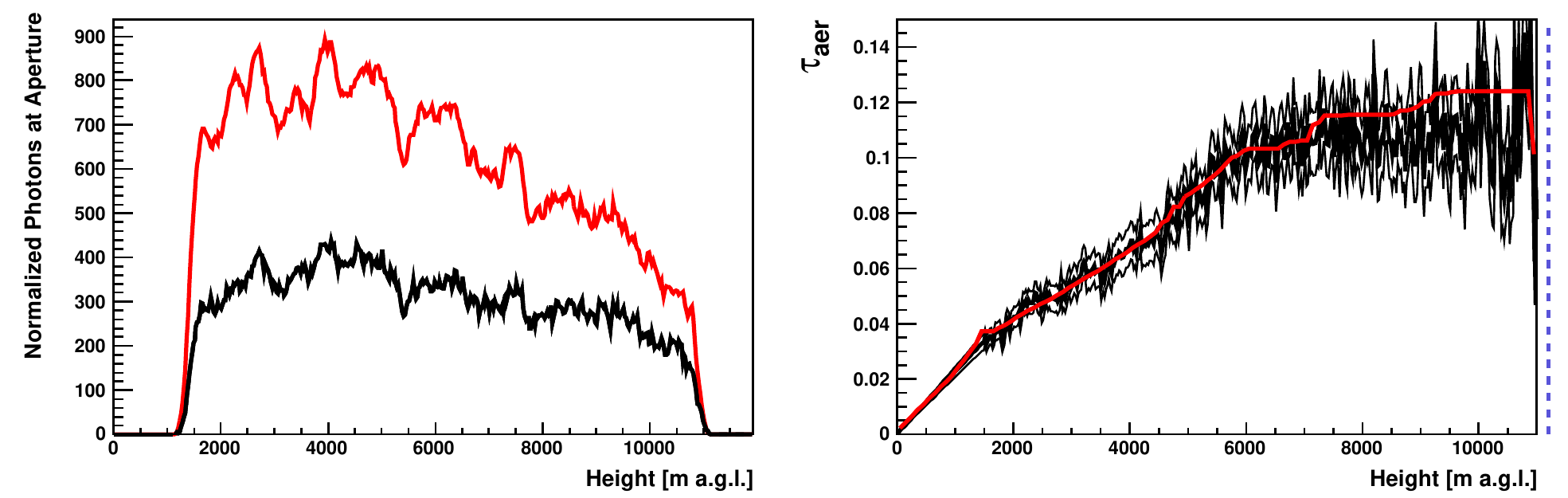}
    \end{minipage}
    \begin{minipage}[t]{.99\textwidth}
      \centering
      \includegraphics*[width=.99\linewidth]{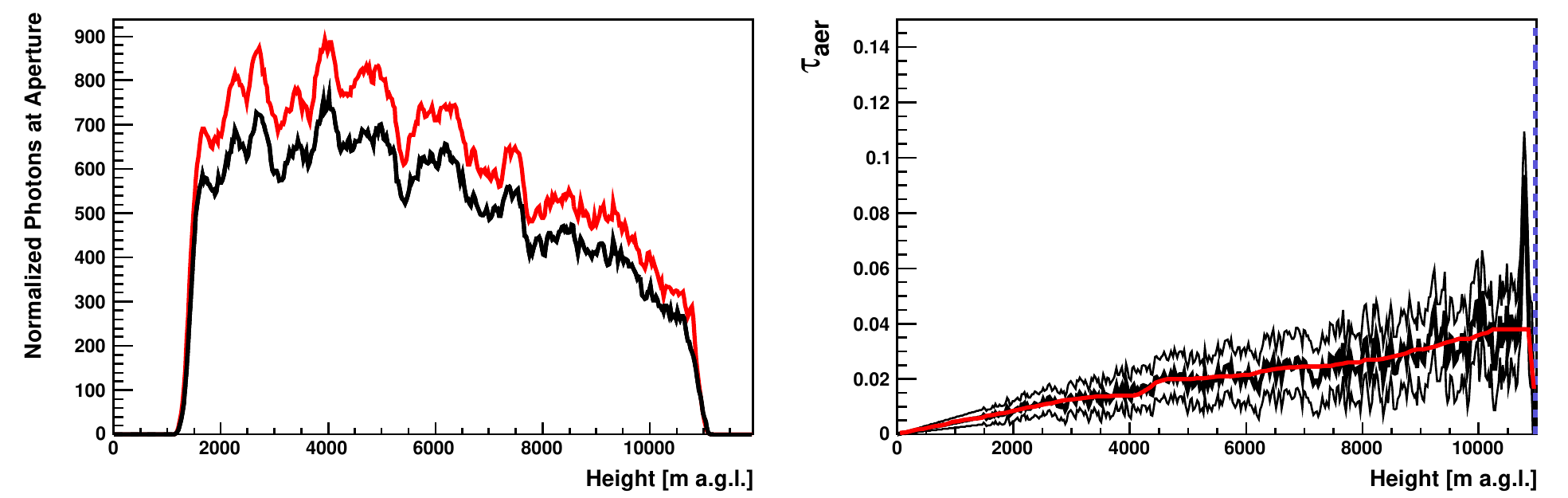}
    \end{minipage}
    \begin{minipage}[t]{.99\textwidth}
      \centering
      \includegraphics*[width=.99\linewidth]{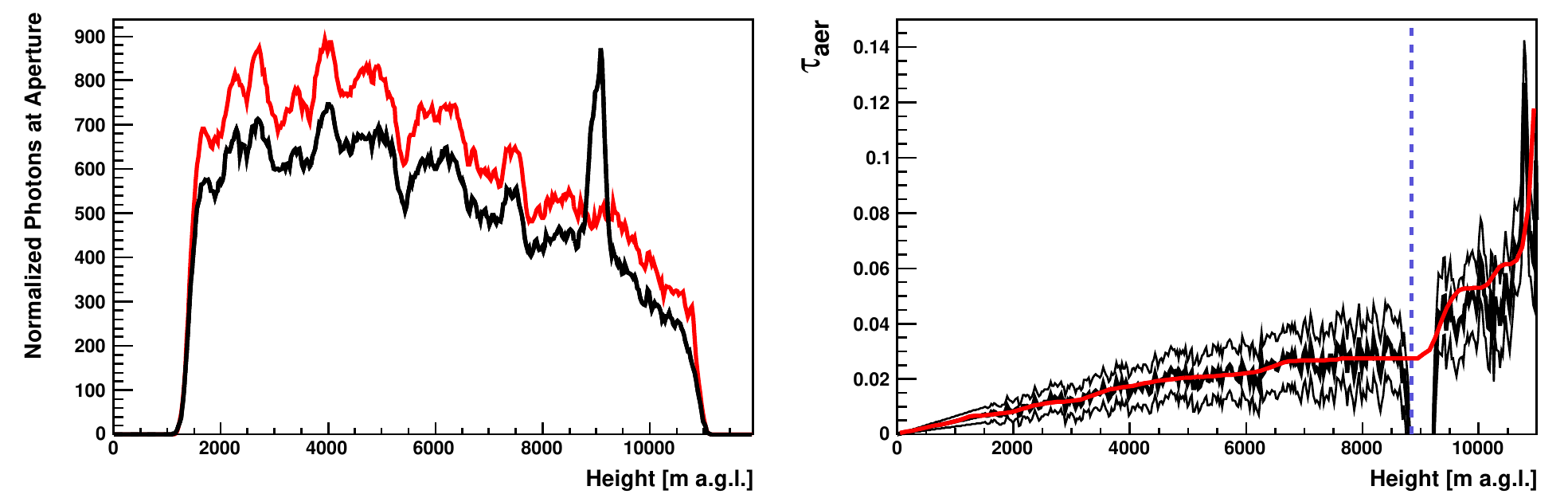}
    \end{minipage}
    \begin{minipage}[t]{.99\textwidth}
      \centering
      \includegraphics*[width=.99\linewidth]{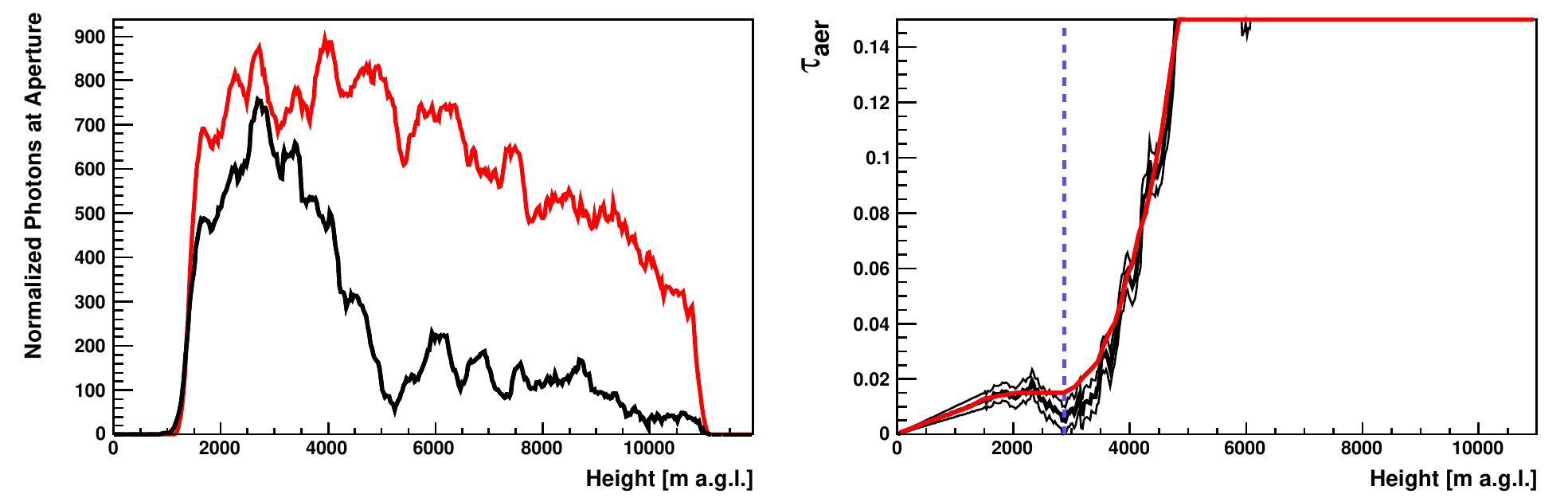}
    \end{minipage}
    \caption[]{\label{fig:amtprofiles}
      From top to bottom: November 11, 2010, 10--11~UTC: Hazy conditions;
      December 1, 2010, 4--5~UTC: Clear conditions; December 1, 2010, 5--6~UTC:
      Laser hit small cloud around 8800\,m a.g.l.; December 7, 2010, 4--5~UTC:
      Cloud obstructing the light above 2800\,m a.g.l. Left: Light profiles as
      measured with the AMT in black, reference clear nights in red. Right:
      Measured \vaod in black with uncertainties, fit \vaod in red. The
      reconstructed minimum cloud height is marked with a blue dashed line.
    }
  \end{center}
\end{figure}

In Fig.~\ref{fig:amtprofiles}, light profiles and aerosol optical depth profiles
for four different conditions are shown. In the left panels, the averaged hourly
light profile is shown together with the reference clear night. The fine
structure of the trace is due to light lost in the gaps between the PMTs of the
camera. On the right side, the measured \vaod profiles are shown as a thick
black line with their uncertainties as thin lines. In red, the fit \vaod
profiles are superimposed, the minimum cloud height is indicated with a blue
dashed line. From top to bottom, hazy\footnote{The measurement of the example
hazy profile was done before the safety daemon on the AMT was configured to
prevent the door from opening after 10~UTC, see Sec.~\ref{sec:slowcontrol}} (a)
and clear conditions (b), as well as two cloud-affected profiles are shown. It
should be noted, that the clear profile (b) and the profile where the laser hit
a small cloud (c) are only separated by one hour, demonstrating the high
variability of the aerosol conditions and the need for hourly aerosol profiles.

The AMT cloud heights agree with the heights determined from the Raman data,
only a few profiles are not marked with a cloud by the Raman \lidar where very
low clouds are found in the AMT data and vice versa. The Raman \lidar can only
detect clouds directly above the laser facility, while the AMT is also affected
by clouds in the path between the laser and the detector.

\begin{figure}[t]
  \begin{center}
    \centering
    \includegraphics*[width=.99\linewidth]{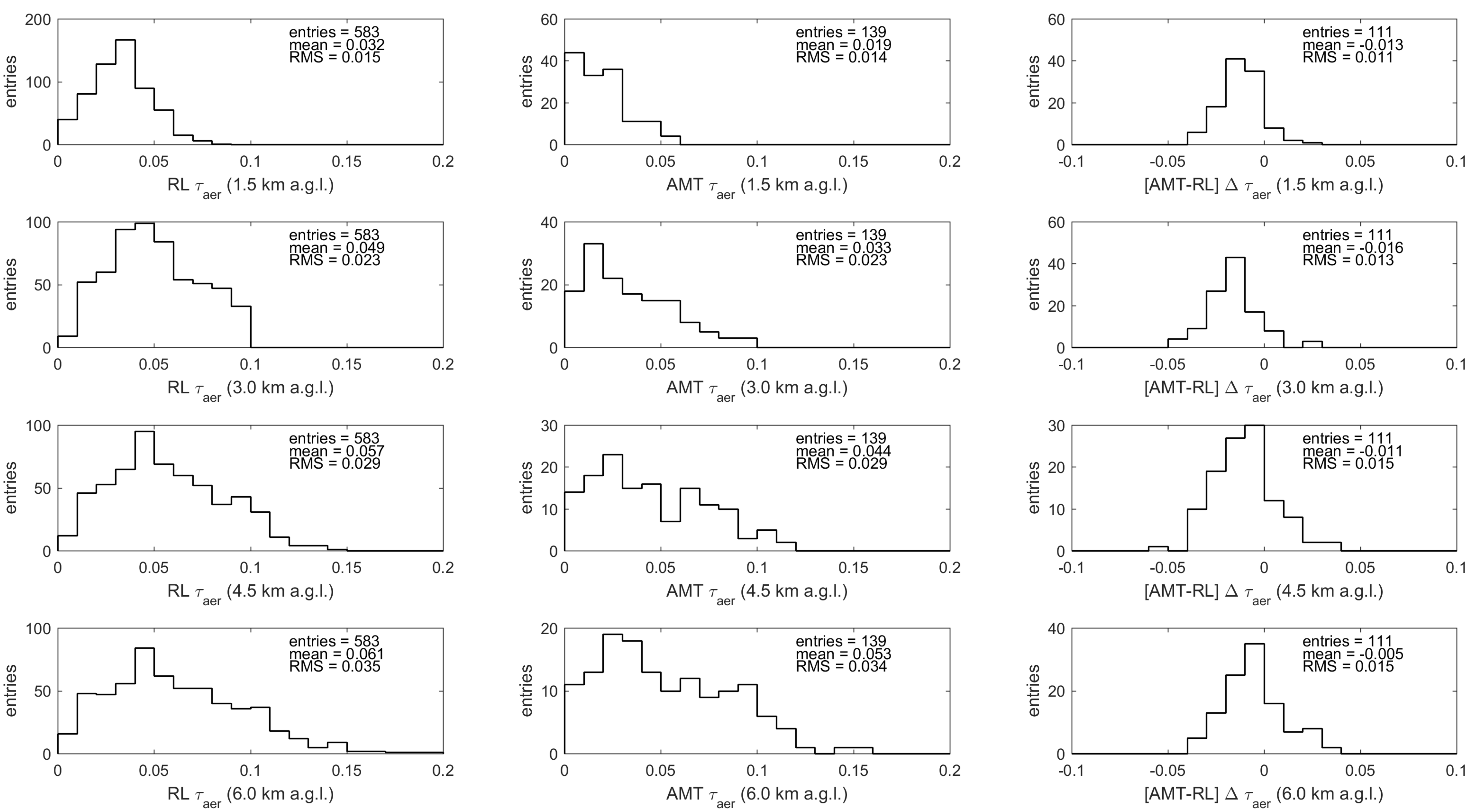}
    \caption[]{\label{fig:vaodhist}
      Mean \vaod and difference in \vaod at 1.5, 3.0, 4.5 and 6.0\,km a.g.l.
      (from upper to lower plot). Left plots: as measured by the Raman \lidar.
      Middle plots: as measured with the AMT. Right plots: the differences in
      \vaod between the two systems.
    }
  \end{center}
\end{figure}

The Raman \lidar \vaod distributions are shown in the left panels of
Fig.~\ref{fig:vaodhist}. A quality cut has been applied to remove the clouds,
the \vaod has to be less than 0.1, 0.15 and 0.2 at 3, 4.5 and 6\,km,
respectively. The mean \vaod at 1.5\,km a.g.l.\ is 0.032 with an RMS of 0.015,
0.049 with an RMS of 0.023 at 3\,km, 0.057 with an RMS of 0.029 at 4.5\,km, and
at 6\,km, the mean is 0.061 with an RMS of 0.035.

The distribution of available AMT \vaod profiles after the selection of the
cases without clouds, using the same criteria as for the Raman \lidar data, is
shown in the middle panels of Fig.~\ref{fig:vaodhist}. The mean \vaod at 1.5\,km
a.g.l.\ is 0.019 with an RMS 0.014, 0.033 with an RMS of 0.023 at 3\,km, 0.044
with an RMS of 0.029 at 4.5\,km, and at 6\,km, a mean \vaod of 0.053 with an RMS
of 0.034 is found.

As expected, both the mean \vaod and the spread increases with height. In the
right panels of Fig.~\ref{fig:vaodhist}, the binned differences in \vaod between
the two analyses are shown for 1.5, 3, 4.5 and 6\,km. Mean differences of
$-$0.013 with an RMS of 0.011, $-$0.016 with an RMS of 0.013, $-$0.011 with an
RMS of 0.015, and $-$0.005 with an RMS of 0.015 are found, respectively. The
differences in \vaod between the two systems is discussed in the next section.

\section{Results
\label{sec:results}}

For the first time, the side-scatter method to obtain vertical aerosol depth
profiles can be directly compared with a Raman \lidar. The differences of \vaod
measurements for periods when both systems recorded data are presented at 1.5,
3, 4.5 and 6\,km above ground.

\begin{figure}[htbp]
  \begin{minipage}[t]{.41\textwidth}
    \centering
    \includegraphics*[width=.82\linewidth]{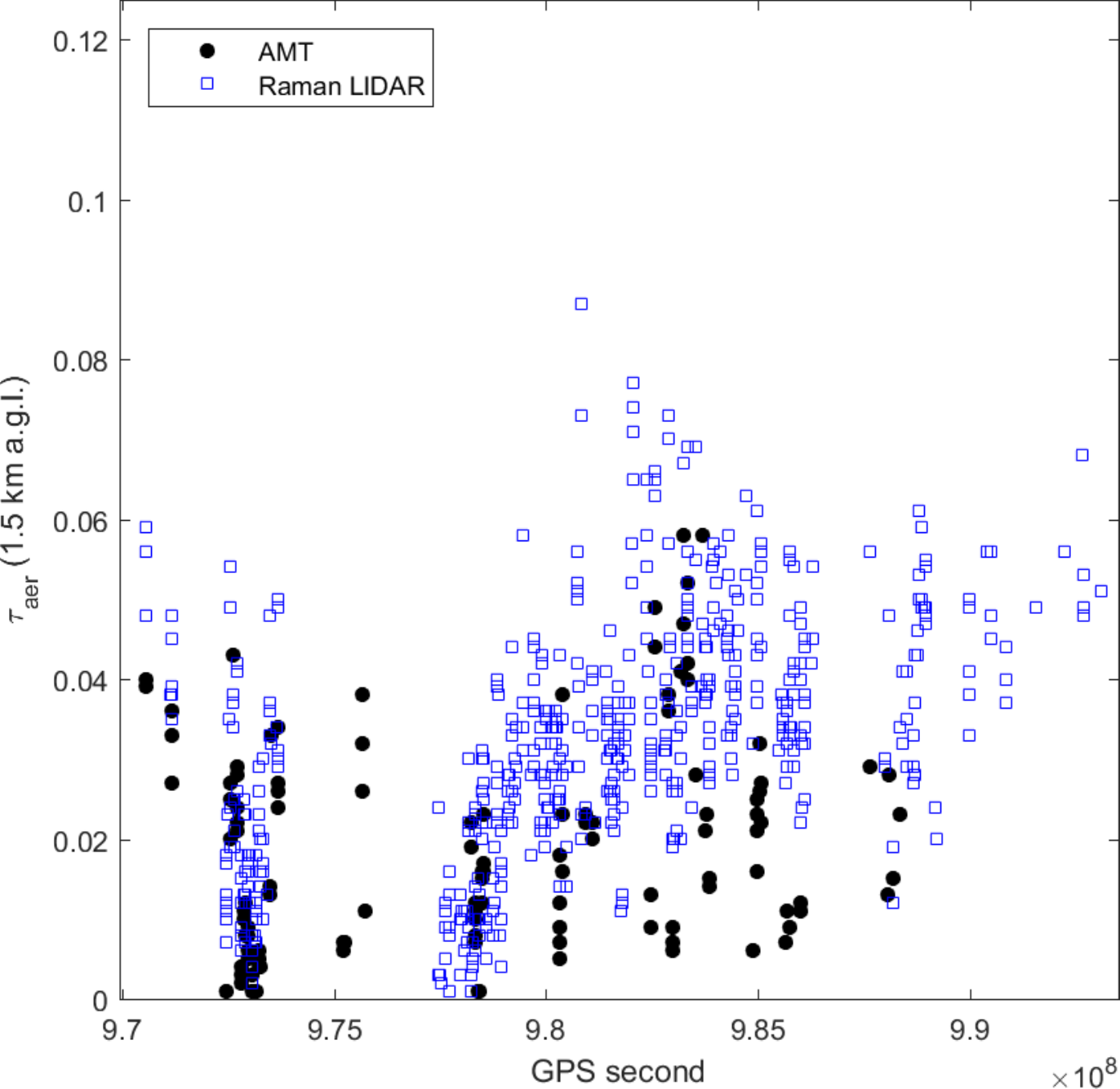}
  \end{minipage}
  \hfill
  \begin{minipage}[t]{.41\textwidth}
    \centering
    \includegraphics*[width=.82\linewidth]{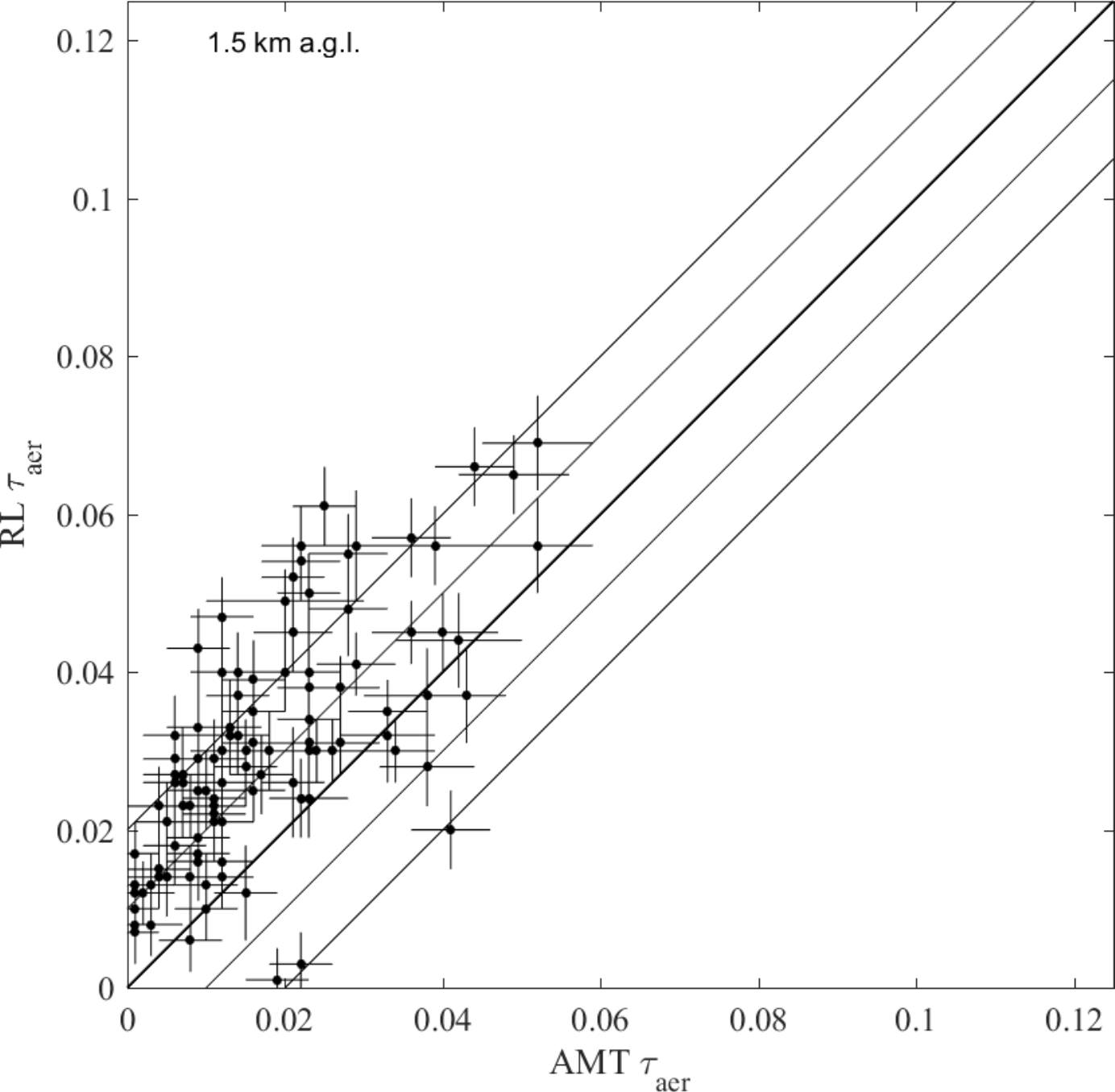}
  \end{minipage}

  \begin{minipage}[t]{.41\textwidth}
    \centering
    \includegraphics*[width=.82\linewidth]{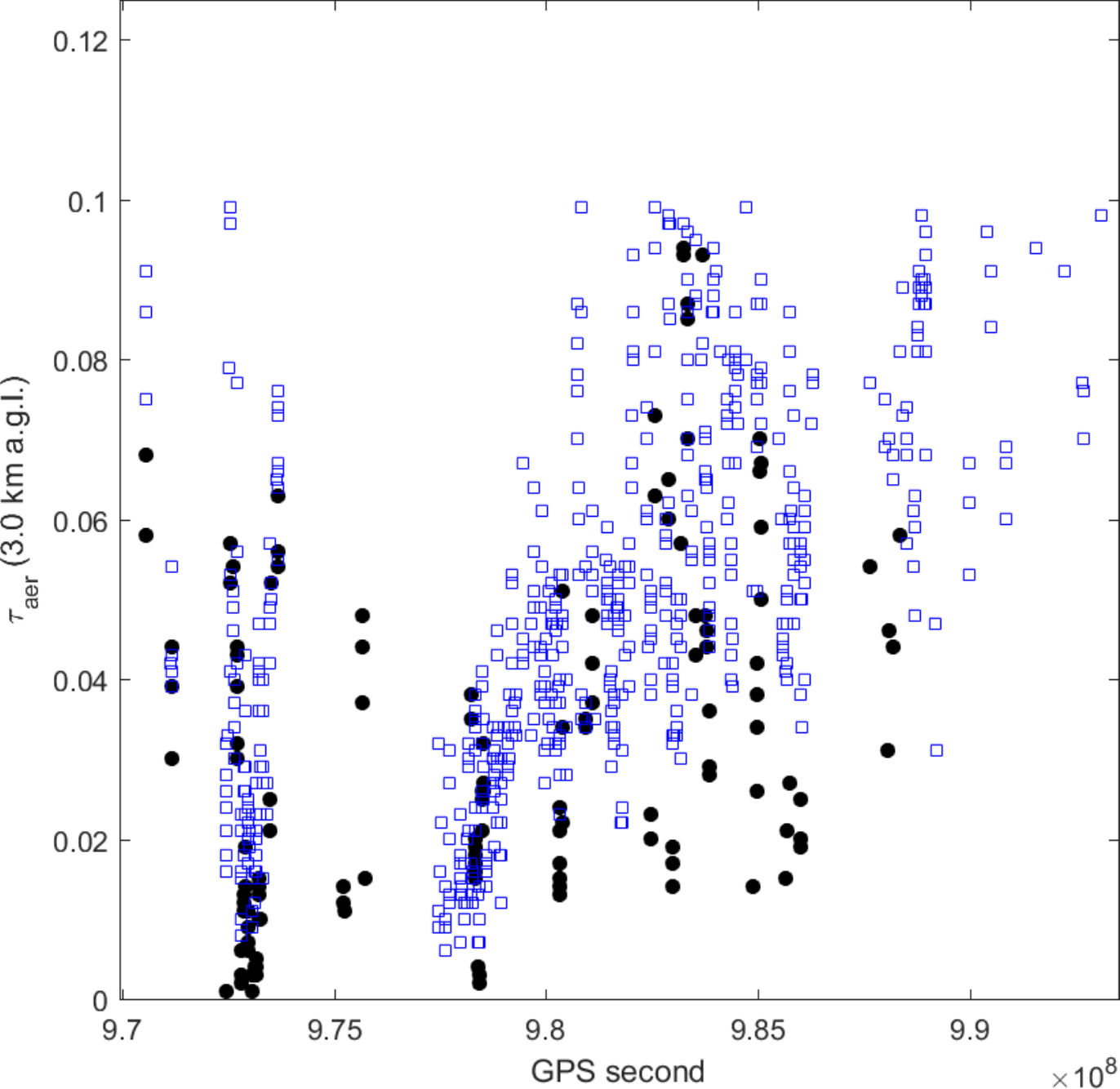}
  \end{minipage}
  \hfill
  \begin{minipage}[t]{.41\textwidth}
    \centering
    \includegraphics*[width=.82\linewidth]{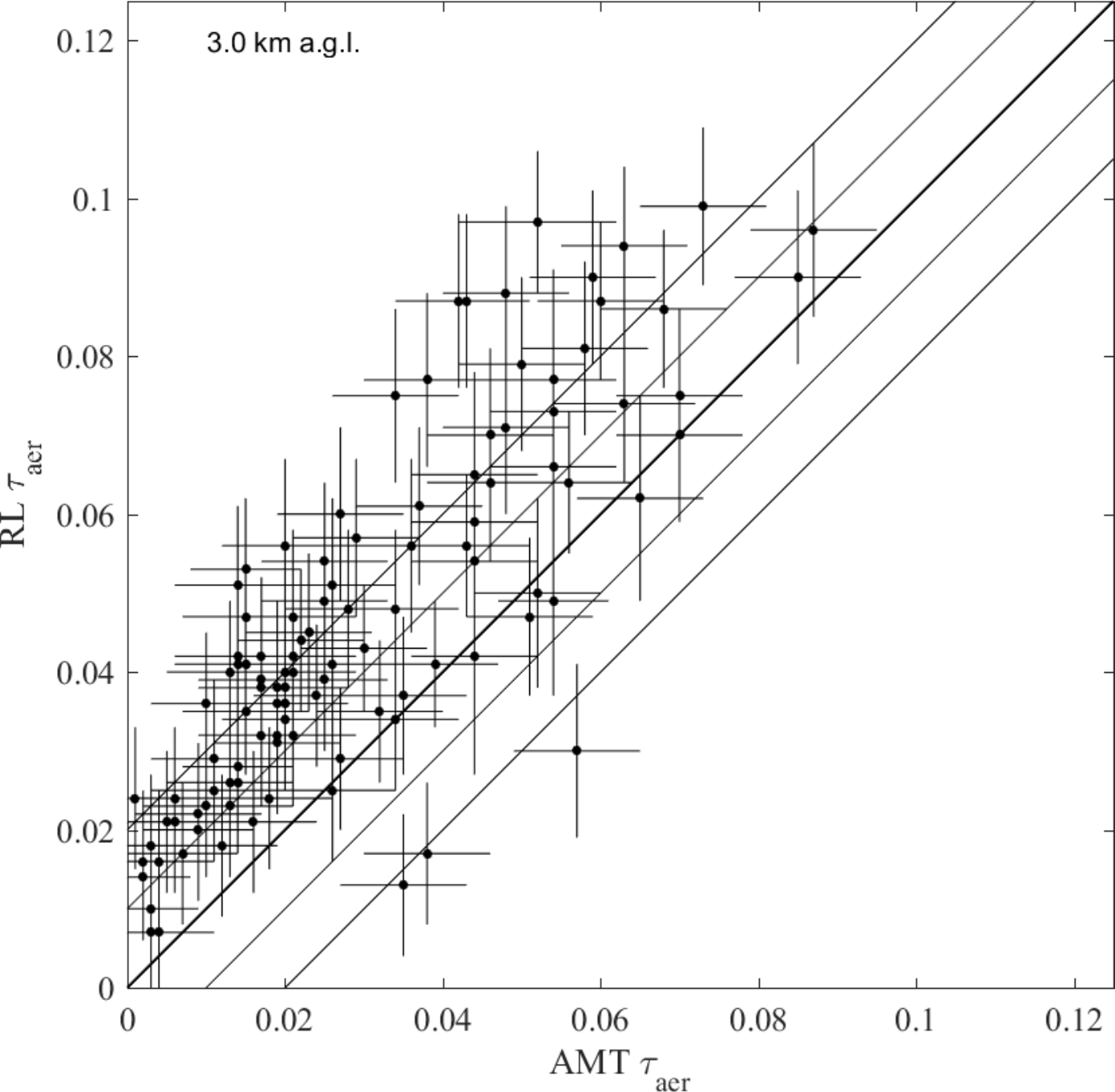}
  \end{minipage}

  \begin{minipage}[t]{.41\textwidth}
    \centering
    \includegraphics*[width=.82\linewidth]{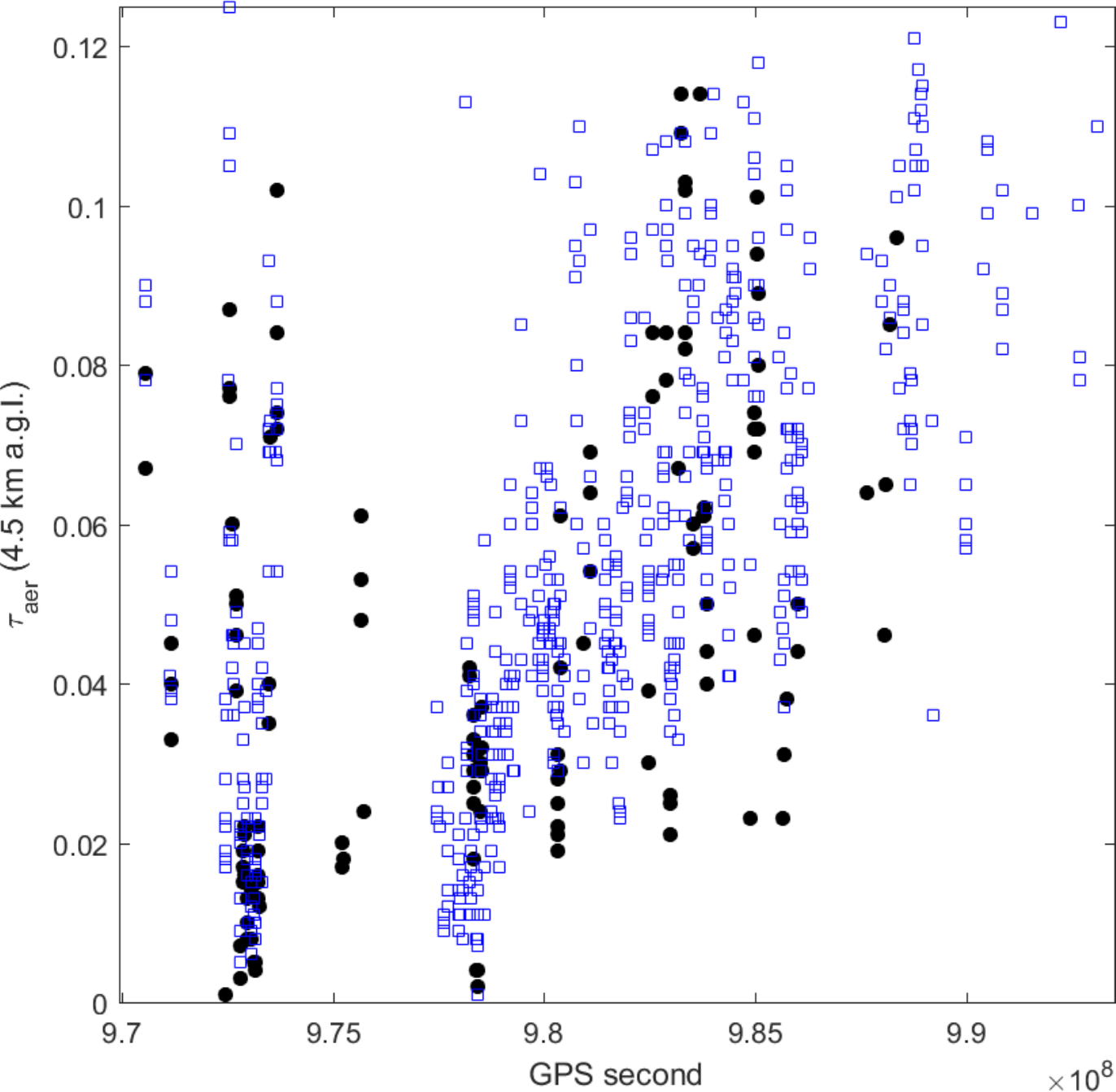}
  \end{minipage}
  \hfill
  \begin{minipage}[t]{.41\textwidth}
    \centering
    \includegraphics*[width=.82\linewidth]{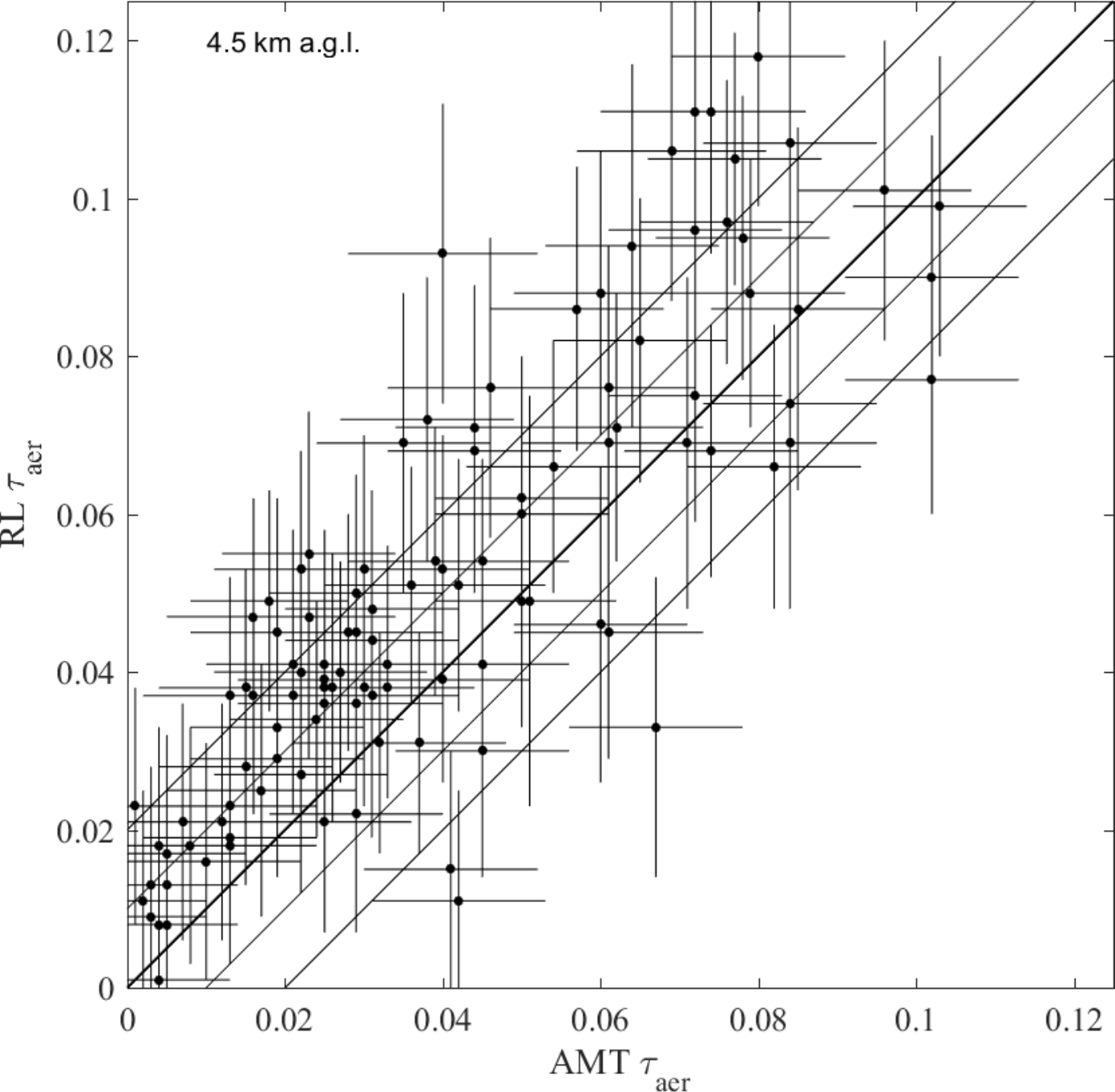}
  \end{minipage}

  \begin{minipage}[t]{.41\textwidth}
    \centering
    \includegraphics*[width=.82\linewidth]{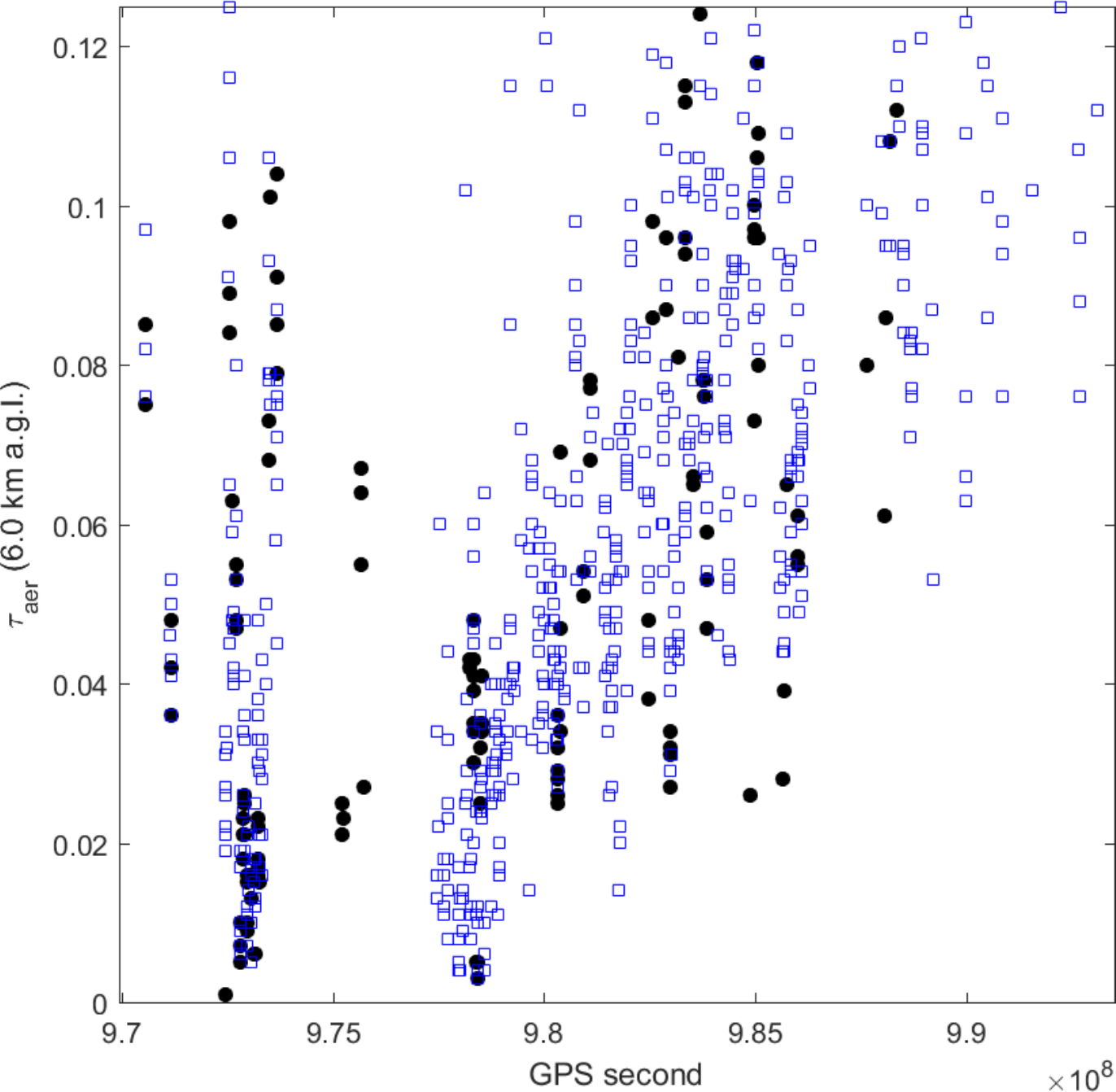}
  \end{minipage}
  \hfill
  \begin{minipage}[t]{.41\textwidth}
    \centering
    \includegraphics*[width=.82\linewidth]{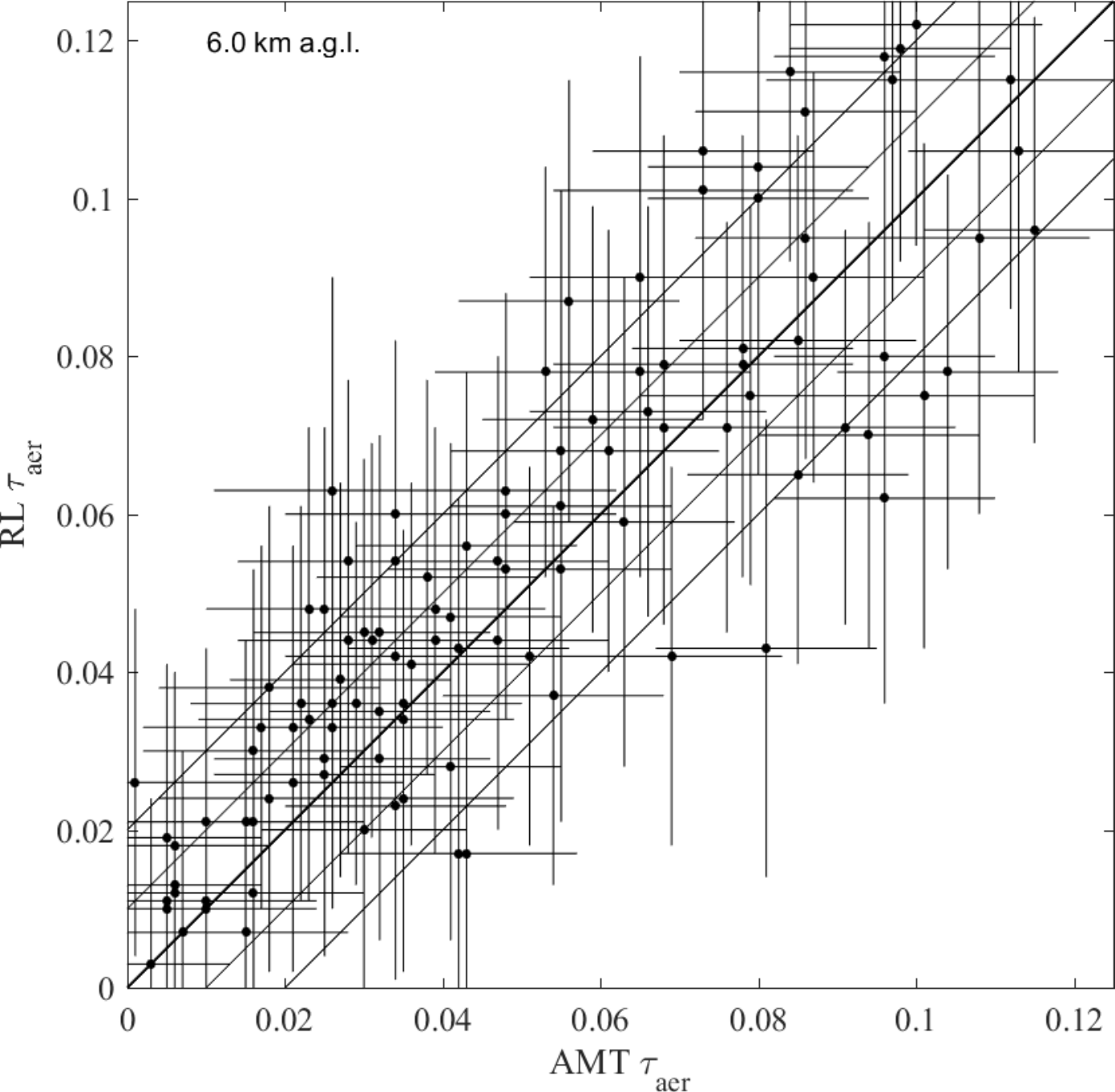}
  \end{minipage}

  \caption[]{\label{fig:vaoddiff}
    Left: \vaod versus time (in GPS seconds) as measured with the AMT (black
    dots) and the Raman \lidar (blue squares) at 1.5, 3, 4.5 and 6\,km a.g.l.\
    between October 1, 2010 and June 30, 2011. Right: \vaod measured with the
    AMT versus Raman \lidar for hours observed with both instruments. The
    diagonal is indicated with the thick black line, the parallel thin black
    lines denote a shift in \vaod of $\pm$0.01 and $\pm$0.02.
  }
\end{figure}

The difference between the \vaod measurements of the AMT and of the Raman \lidar
at various heights is shown in Fig.~\ref{fig:vaoddiff}. In the left column, all
data are shown versus time for 1.5, 3, 4.5 and 6\,km a.g.l. A clear seasonal
trend is visible for both data sets with higher aerosol load in atmosphere
during summer. In the right column of Fig.~\ref{fig:vaoddiff}, the measurements
of \vaod at the Raman \lidar are shown versus the AMT data for common
measurement periods. At 1.5\,km, the Raman \lidar measured \vaod is
systematically higher than the AMT result, as well as at 3, 4.5 and at 6\,km.

There are almost no AMT data in May and in June when the Raman \lidar has
measured a largely increasing \vaod compared to the previous months, so
comparisons in atmospheric conditions with a relatively higher aerosol load are
not available.

The Data Normalization method used to retrieve \vaod from AMT observations needs
a so-called \textit{reference night}, a measurement in which the atmospheric
aerosol content can be considered negligible. The chosen reference
\textit{clean} measurement was on January~7, 2011 between 6 and 8~UTC. The
coincident Raman \lidar measurement of \vaod shows that the aerosol content is
extremely low, but \vaod is about 0.01 above 3~km (upper-left panel of
Fig.~\ref{fig:ramprofiles}), and this value is close to the mean differences
between Raman \lidar and AMT data (right panels of Fig.~\ref{fig:vaodhist}).

It is possible that the \vaod profiles measured by the two systems can also show
a difference in the vertical distribution of the aerosols. The LT and RL are
located close to the city of Lamar at a major highway with truck traffic and is
surrounded by rangeland, the AMT overlooks planted fields and is far away from
both the highway and any kind of civilization. This might cause a difference in
the aerosol type and concentration between the AMT and Raman sites. The AMT
measurements are dominated by the transmission, not the scattering out of the
laser. The difference in surroundings could explain part of the differences and
could also be a source of seasonal variations.

\section{Conclusion
\label{sec:conclusion}}

The Raman \lidar and AMT detector described in this paper were set up in the
field near Lamar, Colorado, USA. Both systems were operated remotely for a
period of 11~months. A comparison between between the elastic side-scatter and
Raman back-scatter methods of aerosol optical depth was obtained. To our
knowledge this is the first time that such a comparison has been made
systematically. A correlation between the measurements was observed. A
systematic offset was also observed. The Raman system obtained higher values of
aerosol optical depth than did the elastic side-scatter system. The most likely
reason for this offset is that the nominally clear ``reference'' nights that
were used to normalize the elastic side-scatter data were not aerosol free.
Systematic horizontal non-uniformity of the atmosphere clarity between the two
instruments may also have contributed to the observed bias. The Raman system was
located in a ranching area relatively close to the small city of Lamar, the
side-scatter detector was located in a farmed area with many fields.

The Raman \lidar system described in this paper has now been relocated to the
central laser facility (CLF) of the Pierre Auger Cosmic Ray Observatory in
Mendoza province, Argentina. A more detailed atmospheric measurement program is
in progress. This program uses the Raman backscatter receiver located at the CLF
and optical fluorescence telescopes located at four sites on the perimeter of
the 3000\,km$^2$ observatory.


\section*{Acknowledgments}

We gratefully acknowledge the valuable assistance provided by the Prowers County
(Colorado) Commissioners and by Bacca County (Colorado) Commissioner Spike
Ausmus. We also acknowledge the manager of Columbia University Nevis Labs, Anne
Therrien, who arranged to make the AMT optics and shelter available for this
project. The Global Data Assimilation System (GDAS) data are taken from the NOAA
Air Resource Laboratory. Part of this work is supported by the
\emph{Bundesministerium f\"ur Bildung und Forschung} (BMBF) under contract
05A08VK1 and by the National Science Foundation under grant 0855680. The
\emph{Center of Excellence CETEMPS/DSFC -- University of L'Aquila} is gratefully
acknowledged.

\bibliographystyle{JHEP}
\bibliography{ColoAtmoRnD}

\providecommand{\href}[2]{#2}\begingroup\raggedright\begin{thebibliography}{10}

\bibitem{OSM}
{OpenStreetMap contributors}, {\it {OpenStreetMap}},  2014.

\bibitem{Abraham:2009fd}
{\bf The Pierre Auger} Collaboration, J.~Abraham et~al., {\it {The Fluorescence
  Detector of the Pierre Auger Observatory}},  {\em Nucl. Instr. Meth.} {\bf
  A620} (2010) 227--251, [\href{http://xxx.lanl.gov/abs/0907.4282}{{\tt
  arXiv:0907.4282}}].

\bibitem{Abbasi:2006aer}
{\bf The High Resolution Fly's Eye (HiRes)} Collaboration, R.~U. Abbasi et~al.,
  {\it {Techniques for measuring atmospheric aerosols at the High Resolution
  Fly's Eye experiment}},  {\em Astropart. Phys.} {\bf 25} (2006) 74--83,
  [\href{http://xxx.lanl.gov/abs/astro-ph/0512423}{{\tt astro-ph/0512423}}].

\bibitem{Abreu:2013clf}
{\bf The Pierre Auger} Collaboration, P.~Abreu et~al., {\it {Techniques for
  measuring aerosol attenuation using the Central Laser Facility at the Pierre
  Auger Observatory}},  {\em JINST} {\bf 8} (2013) 04009,
  [\href{http://xxx.lanl.gov/abs/1303.5576}{{\tt arXiv:1303.5576}}].

\bibitem{Abraham:2010atmo}
{\bf The Pierre Auger} Collaboration, J.~Abraham et~al., {\it {A Study of the
  Effect of Molecular and Aerosol Conditions in the Atmosphere on Air
  Fluorescence Measurements at the Pierre Auger Observatory}},  {\em Astropart.
  Phys.} {\bf 33} (2010) 108--129,
  [\href{http://xxx.lanl.gov/abs/1002.0366}{{\tt arXiv:1002.0366}}].

\bibitem{Fick:2006}
B.~Fick et~al., {\it {The Central Laser Facility at the Pierre Auger
  Observatory}},  {\em JINST} {\bf 1} (2006) 11003.

\bibitem{Pappalardo}
G.~Pappalardo et~al., {\it {Aerosol LIDAR Intercomparison in the Framework of
  the EARLINET Project 3: Raman LIDAR Algorithm for Aerosol Extinction,
  Backscatter, and LIDAR Ratio}},  {\em Appl. Optics} {\bf 43} (2004) 537--538.

\bibitem{Mueller:Raman}
D.~M{\"u}ller, A.~Ansmann, I.~Mattis, M.~Tesche, U.~Wandinger, D.~Althausen,
  and G.~Pisani, {\it {Aerosol-type-dependent lidar ratios observed with Raman
  lidar}},  {\em J. Geophys. Res.} {\bf 112} (2007) D16202.

\bibitem{GDASinformation}
{NOAA Air Resources Laboratory (ARL)}, {\it {Global Data Assimilation System
  (GDAS1) Archive Information}},  {Tech. rep.}, 2004.

\bibitem{Bockmann}
C.~Bockmann et~al., {\it {Aerosol LIDAR Intercomparison in the Framework of the
  EARLINET Project 2: Aerosol Backscatter Algorithms}},  {\em Appl. Optics}
  {\bf 43} (2004) 977--989.

\bibitem{Kokhanovsky:1998}
A.~A. Kokhanovsky, {\it {Variability of the phase function of atmospheric
  aerosols at large scattering angles}},  {\em J. of Atm. Sci.} {\bf 55} (1998)
  314--320.

\bibitem{MacLeod}
H.~A. MacLeod, {\em {Thin-Film Optical Filters}}.
\newblock Taylor \& Francis, 4~ed., 2010.

\bibitem{Wandinger}
U.~Wandinger, {\em {Raman LIDAR}}, vol.~102, pp.~241--271.
\newblock Springer, 233 Spring St., New York, NY 10013, United States, {LIDAR:
  Range Resolved Optical Remote Sensing of the Atmosphere, Springer Series in
  Optical Sciences}~ed., 2004.

\bibitem{Chiao-Yao}
S.~Chiao-Yao, {\it {Spectral structure of laser light scattering revised:
  bandwindths of nonresonant scattering LIDARs}},  {\em Appl. Optics} {\bf 40}
  (2001) 4875--4884.

\bibitem{Whiteman}
D.~N. Whiteman, {\it {Examination of the traditional Raman LIDAR technique. I
  Evaluating the temperature-dependent LIDAR equations}},  {\em Appl. Optics}
  {\bf 42} (2003) 2571--2592.

\bibitem{patent1}
B.~J. Kross, S.~Majewski, C.~J. Zorn, and L.~A. Majewski, {\it {Flexible liquid
  core light guide with focusing and light shaping attachments}},  1997.
\newblock {US Patent \#5684908}.

\bibitem{patent2}
G.~Nath, {\it {Lightguide filled with a liquid containing dimethylsulfoxide}},
  1999.
\newblock {US Patent \#5857052}.

\bibitem{patent3}
G.~Nath, {\it {Flexible lightguide with a liquid core}},  2001.
\newblock {US Patent \#6314226}.

\bibitem{Boyer:HiRes}
J.~Boyer, B.~Knapp, E.~Mannel, and M.~Seman, {\it {FADC based DAQ for HiRes
  Fly's Eye}},  {\em Nucl. Instr. Meth.} {\bf A482} (2002) 457--474.

\bibitem{Wiencke:1999}
{\bf The High Resolution Fly's Eye (HiRes)} Collaboration, L.~Wiencke et~al.,
  {\it {Radio-controlled Xenon Flashers for Atmospheric Monitoring at the HiRes
  Cosmic Ray Observatory}},  {\em Nucl. Instr. Meth.} {\bf A428} (1999)
  593--607.

\bibitem{Brack:Cal}
J.~T. Brack, R.~Cope, A.~Dorofeev, B.~Gookin, J.~L. Harton, Y.~Petrov, and
  A.~C. Rovero, {\it {Absolute calibration of a large-diameter light source}},
  {\em JINST} {\bf 8} (2013), no.~05 P05014,
  [\href{http://xxx.lanl.gov/abs/1305.1329}{{\tt arXiv:1305.1329}}].

\bibitem{Mathes:2011icrc}
{\bf The Pierre Auger} Collaboration, H.~J. Mathes et~al., {\it {The HEAT
  telescopes of the Pierre Auger Observatory: status and first data}},  in {\em
  Proc. 32nd ICRC}, vol.~3, (Beijing, China), pp.~149--152, 2011.
\newblock \href{http://xxx.lanl.gov/abs/1107.4807}{{\tt arXiv:1107.4807}}.

\bibitem{Abraham:2009pm}
{\bf The Pierre Auger} Collaboration, J.~Abraham et~al., {\it {The Fluorescence
  Detector of the Pierre Auger Observatory}},  {\em Nucl. Instr. Meth.} {\bf
  A620} (2010) 227--251, [\href{http://xxx.lanl.gov/abs/0907.4282}{{\tt
  arXiv:0907.4282}}].

\end{thebibliography}\endgroup

\clearpage

\end{document}